\documentclass[twocolumn]{aa}
\usepackage{graphicx}
\usepackage{epsf}
\usepackage{rotate}
\usepackage{rotating}

\usepackage{psbox}

\begin{document}
   \title{The Near-Infrared properties of the host galaxies of radio quasars }

%   \subtitle{I. Overviewing the $\kappa$-mechanism}

   \author{S.F.S\'anchez\thanks{Visiting Astronomer, German-Spanish Astronomical Centre, Calar Alto, operated by the Max-Planck-Institute for Astronomy, Heidelberg, jointly with the Spanish National Commission for Astronomy.}
          \inst{1}
          \and
          J.I. Gonz\'alez-Serrano
          \inst{2}
          }

   \offprints{S.F.S\'anchez}

   \institute{Astrophysikalisches Institut Potsdam (AIP), An der Sternwarte 16, D-14482 Potsdam, Germany\\
              \email{ssanchez@aip.de}
         \and
                Instituto de F\'\i sica de Cantabria (IFCA), Universidad de Cantabria-CSIC, Avd. de Los Castros S/N, 35005-Santander, Spain
             \email{gserrano@ifca.unican.es}
             }

   \date{Received July 26th, 2002; accepted April 4th, 2003}

   \abstract{
     
     We present $K$-band images of 31 radio quasars selected from the
     B3-VLA quasar sample.  A new method has been developed to detect
     and restore the host galaxies of these quasars.  We have detected
     the host galaxies of 16 of them. Using these data together with
     previous data from the B3-VLA quasars (Carballo et al. 1998,
     hereafter Paper I) and literature data from other samples, we
     have analyzed a sample of 69 hosts of radio quasars, covering a
     redshift range between 0$<\sim z<$3.
     
     The host galaxies are large ($r_{\rm e}\sim$15kpc) and luminous
     elliptical galaxies ($\sim$75\% of them brighter than $L_{K}^*$),
     with an evolution similar to that of radio galaxies.  A
     significant fraction ($\sim$40\%) of them shows evidence of a
     possible collision/merging process. They follow a $\mu_{\rm
     e}$$-$$r_{\rm e}$ relation similar to that of normal elliptical
     galaxies. The morphological and photometric similarities between
     these galaxies and radio galaxies in this wide range of redshifts
     is a good test of the reliability of unification schemes.  All of
     them show little evolution from $z$$=$3 to the present epoch.
     Their $K-z$ distribution is consistent with a no-evolution model,
     with a fraction of the dispersion due to differences in radio
     power.  We have found a correlation between the quasar radio
     power and the host luminosity.  We have also found a correlation
     between the host and nuclear source luminosities.  These
     correlations could be induced by a physical relation between the
     central black hole and the bulge mass (Magorrian et al. 1998).

   \keywords{Galaxies: quasars: general --
               }
   }

%\titlerunning
   \maketitle

\section{Introduction}

The studies of the host galaxies (HGs) of active galactic nuclei
(AGNs) can help us to answer important questions related with these
peculiar objects: (i) what is the origin of the nuclear activity? (ii)
what  is the origin of radio emission? (Smith \& Heckman 1990;
Hutchings \& Neff 1992), (iii) which AGNs can be unified by
orientation effects (standard unification schemes, Antonucci 1993;
Urry \& Padovani 1995) and which not? (e.g., Ellingson et al. 1991),
(iv) which is the proper model to explain the quasars evolution?
(e.g., Kauffmann \& Haehnelt 2000). It is necessary to determine the
space of parameters filled by the different families of AGNs in order
to determine which apparently normal galaxies could contain a dormant
AGN.

\begin{table*}
\begin{center}
\caption{Characteristics of the data}
\label{tab:cali}
\vspace{0.2cm}
\begin{tabular}{@{}rccrccrcc}
\hline
Date &Telescope&Instrument&FOV&Pixel Scale&Exp. time& $\sigma_{K}$ &${\rm seeing}$ &$\mu_{\rm K,lim}$\\
(1)  & (2)     &   (3)    & (4) & (5)           & (6) & (7) & (8)  & (9) \\
\hline
\hline
29 Sep 96 &4.2m WHT&WHIRCAM&$\sim1\arcmin\times1\arcmin$&0.240$\arcsec$/pixel&900s&0.03 &0.8$\arcsec$ &21.8 \\
14 Oct 97 &3.5m CAHA&OMEGA&$\sim6\arcmin\times6\arcmin$&0.396$\arcsec$/pixel&1800s&0.06 &1.2$\arcsec$ &20.9 \\
15 Oct 97 &3.5m CAHA&OMEGA&$\sim6\arcmin\times6\arcmin$&0.396$\arcsec$/pixel&1800s&0.16 &1.1$\arcsec$ &20.9 \\
16 Oct 97 &3.5m CAHA&OMEGA&$\sim6\arcmin\times6\arcmin$&0.396$\arcsec$/pixel&1800s&0.12 &0.9$\arcsec$ &20.9 \\
3 Feb 99  &3.5m CAHA&OMEGA&$\sim6\arcmin\times6\arcmin$&0.396$\arcsec$/pixel&2700s&0.09 &1.3$\arcsec$ &21.0 \\
\hline
\end{tabular}
\end{center}

(1) Observing date. (2) Observing telescope. (3) NIR instrument used. (4) Field of view of the images. (5) Pixel scale. (6) Exposure time used in the science exposures. (7) Photometric error derived from the standard deviation around the mean photometric zero-point per night. (8) Mean FWHM of field stars on the images along each night. (9) Mean 3$\sigma$ limiting surface brightness magnitude of the science exposures for each night.

\end{table*}

Studies of samples of low luminosity and low redshifts AGNs (e.g.,
MacKenty 1990; Gonz\'alez-Serrano et al. 1993) showed a difference in
the morphology of the host galaxies of radio loud (elliptical) and
radio quiet (spiral) sources. Preliminary studies of the HGs of
quasars from groundbased telescopes apparently supported this
hypothesis to explain radio activity (e.g. Smith et al. 1986;
V\'eron-Cetty \& woltjer 1990). These studies are limited by the
atmospheric seeing: The central point-like source contaminates the
surrounding HG, and/or even blurred by it, due to the point spread
function (PSF) . Due to this, these studies are very complex,
especially at high $z$, where the surface brightness of the HGs drops
significantly due to cosmological dimming. Observing at the
near-infrared (NIR) reduces the central point-like contribution, since
the ratio between the luminosity of the nuclear source and the host is
minimized (Dunlop et al. 1993). This strategy has been used by several
authors (McLeod \& Rieke 1994a,b; Taylor et al. 1996; Hutchings \&
Neff 1997; Kotilainen et al. 1998; Carballo et al. 1998; Aretxaga et
al. 1998; Percival et al. 2001). At low $z$ ($z\le$0.4) they found
that (i) the radio quasars inhabit large and luminous elliptical
galaxies ($r_{\rm e}\sim$10kpc and $L_{\rm HGs}>$2$L^*$), and (ii) a
considerable fraction of the radio-quiet quasars inhabits also
elliptical galaxies, and not only spiral ones, in opposition with what
it was previously claimed. The most luminous quasars inhabit
elliptical rather than spiral galaxies, despite their radio-loudness.
Detailed morphological analysis at higher $z$ ranges was not possible.
It seems that the HGs of different families of radio sources have
similar morphologies (down to $z<$0.4) and luminosities (down to
$z<$1), which reinforces the standard unification schemes of radio
sources.

More detailed morphological studies have been done using the {\it HST}
(e.g. Bahcall et al. 1994, 1995a,b, 1996, 1997; Hutchings et al. 1994;
Disney et al. 1995; Hutchings \& Morris 1995; McLure et al.  1999;
Lehnert 1999a,b; Kirhakos et al. 2001; Hutchings et al. 2002). They
show that there is a large fraction of HGs with unusual and irregular
morphologies, undergoing a merging process or with evidence of such a process
in their recent histories. However, these results were not completely
conclusive due to: (i) the saturation and undersampling of the
point-like source in the {\it HST} deep images (e.g., Hutchings et al.
1995; McLure et al.  1999); (ii) strong contamination from emission
lines in the optical bands at certain redshifts; (iii) the use of
bands that sample the flux down to $\lambda \sim$4000\AA, in a range
not dominated by the stellar emission, which limits the redshift range
for studies that use optical bands (e.g. $z<$0.6 for the $R$-band); (iv)
studies mainly based on low-$z$ samples or with a reduced number of
objects. On the other hand, the few published studies using NICMOS data
appear to confirm the groundbased NIR studies (e.g., McLeod \& McLeod
2001; Kukula et al. 2001; Ridgway et al. 2001).

The number of spectroscopic studies of HGs is still limited due mainly
to the technical difficulties. The first attempts to determine the
nature of the `Fuzz' detected around the quasars were made nearly two
decades ago (Boroson \& Oke 1982, 1984, and references therein). With
a large number of caveats, they found that the `Fuzz' around QSOs has
an stellar nature. More recent studies (Canalizo \& Stockton 2000;
Hughes et al. 2000; Nolan et al. 2001; Canalizo \& Stockton 2001;
Courbin et al. 2002) have confirmed the stellar nature of the extended
emission.  They found that at low $z$ these objects are dominated by
an old stellar population ($t_{\rm age}>$12Gyr), with possible traces
of more recent star formation events that might be due to past
interactions or even major mergers (Canalizo \& Stockton 2001;
Stockton \& Ridgway 2001).

There have been few attempts to study the properties of HGs of radio
loud quasars over a large and representative sample. Lehnert et
al. (1999a) studied a sample of 43 3CR radio quasars, using $R$-band
{\it snapshots} HST images. They detected the host for about $\sim$
50\% of their quasars. These HGs present distorted morphologies, close
companions (in a $\sim$25\% of the objects), and a significant
alignment with radio emission. As we quoted above, $R$-band images do
not trace well the stellar population of the HGs at $z>$0.6. The use
of this band combined with the depth of images (16 of the 19
undetected hosts are at $z>$1), and the lack of morphological
classification make their results inconclusive. In contrast, McLure et
al. (1999) and Dunlop et al. (2001) found, from $R$-band deep HST
imaging of a sample of $z<$0.4 objects, that the morphologies of the
HGs of radio-loud and radio-quiet quasars and of radio galaxies were
very similar (elliptical galaxies), and that there were no significant
traces of interactions and/or distorted morphologies. Their sample
covers a lower range of radio powers than the one studied by Lehnert
et al. (1999a). It might be impossible to generalize the properties of
the HGs of powerful samples of radio quasars, like 3CR, since these
objects trace only the most violent events of the radio source family.

It is necessary to study the HGs of radio quasars in a wide range of
redshifts, for complete samples with intermediate radio power, trying
to trace the stellar component of their HGs. The $K$-band is the best
choice to compare with other families of radio sources since they
have been previous studied in this band (e.g., Lilly et al. 1985;
Willot et al. 2002). We started this study in 1996 using the B3-VLA
quasar sample (Vigotti et al. 1997; S\'anchez et al. 2001). This
sample consists of 130 radio quasars selected up to $S_{\rm
408MHz}>$0.1 Jy. As a part of it we published the results based on the
study of the $K$-band images of 54 quasars of this sample (Carballo et
al. 1998, hereafter Paper I), 32 of them were analyzed in order to
detect the HG and determine its photometry. It was possible to recover
the HG in 16 of them (50\%); These were large and luminous galaxies,
roughly similar to radio galaxies down to $z<$1. We present here the
study of new $K$-band images of 31 radio quasars from the B3-VLA
sample.

The distribution of this paper is as follows: in Sect. 2 we present
the sample and the observations and in Sect. 3 we show the technique
used to restore the HGs of the quasars. In Sect. 4 we present the
results of this analysis. In Sect. 5 we discuss the evolution of HGs
of radio sources, showing its $K-z$ distribution, the structural
parameters of these HGs, and its relation with radio power and nuclear
emission. In Sect. 6 we study the relation between nuclear emission
and radio power. We present the conclusions in Sect. 7.  Throughout
this article we have assumed a standard cosmology with $H_{\rm o}$=50
Km s$^{-1}$ Mpc$^{-1}$ and $q_o$=0.5. The selection of other
cosmological parameters will not significantly affect our results
\footnote{A cosmology with $H_{\rm o}$=65 Km s$^{-1}$ Mpc$^{-1}$,
$\Omega$=0.7 and $\lambda$=0.3 will change the absolute magnitudes of
the objects by $\sim$0.3 mag in the range of redshifts we considered
}.

\section{Observations, data reduction and photometry}

Table \ref{tab:cali} shows a summary of the observations, including
the observing nights, the telescopes and instruments used, their field
of view (FOV) and pixel scale, the exposure time, the error of the
photometric calibration ($\sigma_{K}$, explained below), the mean
seeing during each night (determined from the mean FWHM of the stars
observed along the night) and the mean surface brightness limit at
3$\sigma$ for each night.

A standard NIR observing procedure was used: $n$ unregistered
exposures of $t$ seconds each one were taken and the average image was
registered. A number of $k$ average images were obtained following a
grid pattern around the object central position, with an offset of $l$
arcseconds. Both the target and the photometric standards were
observed using the same procedure. The selection of the
above-mentioned parameters ($n$,$t$,$k$ and $l$) depended on the
characteristics of the detector and on the sky background level of the
night. For the WHIRCAM images 75 unregistered exposures of 2 second,
and 6 average images with a 17$\arcsec$ offset were taken, which
allowed a total exposure time of 900 seconds. For the October 1997
OMEGA images 25 unregistered exposures of 2 seconds, and 9 average
images with a 35$\arcsec$ offset were taken. The procedure was
repeated 4 times to reach a total exposure time of 1800~s. For the
February 1999 images the exposure time of the unregistered frames was
4 seconds and the whole procedure was repeated only three times. The
total exposure time was 2700~s.

The data were reduced using standard IRAF\footnote{IRAF is distributed
by the NOAO, which is operated by AURA, Inc., under contract to the
NSF} packages. First, a dark frame, obtained before each sequence of
exposures along the grid, was subtracted from each registered
image. For the OMEGA images, domeflats were obtained as described in
the OMEGA observer's manual. For the WHIRCAM images a skyflat was
built computing the mean of all the registered images along the night,
applying a 1$\sigma$ rejection algorithm to remove source
contributions.  These flat-field images were then used to correct the
pixel-to-pixel gain variations for each registered image. The sky
background frame was then obtained for each sequence of $k$ images
(six for WHIRCAM, nine for OMEGA), computing the mean of these images,
applying a 1$\sigma$ rejection algorithm to remove source
contributions.  The sky-flux frame was subtracted from each of the
images of the sequence. Once sky-subtracted, the images of each
sequence were re-centred and co-added. When more that one sequence was
obtained for a given object, we have re-centred and co-added all the
sequences to obtain the final image.

All the images were obtained in photometric conditions. Flux
calibration was carried out using UKIRT faint standard stars (Casali
\& Hawarden 1992) observed during each night. The number of
calibration exposures was always more than 12 per night for the OMEGA
runs, and 4 for the WHIRCAM run (which comprised only $\sim$2.5
hours), taken in the course of the night at different air-masses. Due
to the small extinction in the NIR no air mass correction was needed,
as we did not expect significant differences between the derived
calibration zero-points at different air-masses. We used the mean
zero-point for each night in order to calibrate the exposures. The
standard deviation around this value yields the photometric error due
to the calibration for each night. The $K$-band magnitudes were them
obtained by standard aperture photometry, selecting for each object
the aperture where the flux reaches the background.

\section{Analysis of the data}

As we mention in the introduction there are various problems in the
detection and restoration of the HGs of quasars at high $z$; among
these we consider that the most important ones are: (i) the active
nucleus dominates the emission over a wide range of wavelengths, (ii)
the central source flux contaminates the HG flux, due to the width of
the PSF, (iii) the angular sizes of the galaxies are smaller at
$z\sim$1 than at low $z$($<$0.4), viz. in the range of a few
arcseconds, and (iv) at higher redshift, the angular size increases
again, but cosmological dimming makes the surface brightness of the
HGs drop significantly following a $(z+1)^4$-law (i.e., about 3
magnitudes from $z$$\sim$0 to $z$$\sim$1).

We have minimalized as much as possible the effects of these problems.
The first problem was reduced by the selection of the wavelength. At
$\lambda\sim1\mu$m the SED of the galaxies presents a peak, whereas
the SED of the quasars presents a minimum (Dunlop et al.  1993).
Therefore, at $z\sim$1 the ratio between HG and quasar fluxes is
maximal in the $K$-band ($\lambda_{\rm eff}\sim 2 \mu$m). An
additional advantage is that in the NIR the HG emission is dominated
by its stellar component, roughly related with the mass of the
galaxy. This is in contrast with the optical bands, where the
detection of extended emission around the quasars does not guarantee
that the HG has been detected, since there are other more important
contributions to the extended flux, such as gas flow emission (e.g.,
Heckman et al.  1991).

%\begin{table*}
\begin{sidewaystable*}
\begin{center}
\caption {Sample of B3-VLA quasars studied: results of the analysis}
\label{tab:Sample}
\vspace{0.2cm}
\begin{tabular}{lrcrccrrrcrrrrrrrr}
\hline\hline
Object&z&S$_{\rm 408}$ &$\alpha_{\rm 408}^{\rm 1460}$&Obs.&K mag&$\chi^2/\nu$&$P_{\rm ext}$&\%$_{\rm ext}$&Ext.&$\alpha$&$P_{\rm e}$&$P_{\rm d}$& $P_{\rm e}/P_{\rm d}$ &$K_{\rm HG}$&$K_{\rm nuc}$&$r_{\rm e}$\\
 (1)& (2)& (3) & (4) & (5) & (6) & (7) & (8) & (9) & (10) & (11) & (12) & (13) & (14) & (15) & (16) & (17) \\
\hline
0006+397 &1.830&  1.15&  0.66&$O_1$&16.72$\pm$0.12&   1.98&95.8&19.5&Yes&0.55&48&13&3.7&18.5$\pm$0.2&17.0$\pm$0.2&2.39$\pm$0.09\\
0019+431 &1.050&  2.22&  1.34&$O_1$&15.47$\pm$0.12&   0.38&90.4& 5.5&No&$-$&$-$&$-$&$-$&$-$&$-$&$-$ \\
0022+390 &1.932&  1.10&  0.10&$O_1$&16.65$\pm$0.13&   8.70&99.8&41.4&Yes&0.38&84& 9&9.1&17.6$\pm$0.2&17.2$\pm$0.1&2.00$\pm$0.02\\
0032+423 &1.588&  0.92&  0.94&$W$       &16.24$\pm$0.08&$<$0.03&90.0&$<$1.0&No&$-$&$-$&$-$&$-$&$-$&$-$&$-$ \\
0110+401 &1.479&  1.08&  0.65&$O_1$&17.13$\pm$0.09&   4.48&98.6&51.8&Yes&0.40&84&11&7.7&17.9$\pm$0.2&17.9$\pm$0.2&3.60$\pm$0.02\\
0137+401 &1.620&  0.82&  0.85&$O_1$&16.94$\pm$0.05&   0.91&91.8&16.5&No&$-$&$-$&$-$&$-$&$-$&$-$&$-$ \\
0143+446 &0.813&  1.71&  0.99&$O_1$&13.99$\pm$0.16&   6.89&99.5&11.5&?&0.40&84&11&7.7&16.3$\pm$0.1&14.1$\pm$0.2&7.09$\pm$0.02\\
0144+430 &1.790&  0.92&  1.01&$W$       &16.17$\pm$0.03&$<$0.03&90.0&15.3&No&$-$&$-$&$-$&$-$&$-$&$-$&$-$ \\
0144+432 &1.260&  0.82&  0.70&$O_1$&17.12$\pm$0.18&   8.73&99.8&80.2&Yes&0.40&84&11&7.7&17.4$\pm$0.1&18.9$\pm$0.4&4.45$\pm$0.02\\
0157+442 &0.721&  3.26&  0.91&$O_1$&16.32$\pm$0.17&  18.74&99.9&83.4&Yes&0.39&84&11&7.7&16.5$\pm$0.1&18.3$\pm$0.4&4.09$\pm$0.02\\
0158+394 &0.780&  0.12&  0.18&$O_2$&16.34$\pm$0.07&   5.45&99.1& 6.8&No&$-$&$-$&$-$&$-$&$-$&$-$&$-$ \\
0217+417 &1.430&  0.85&  0.94&$O_1$&17.18$\pm$0.15&   2.16&96.2&$<$1.0&No&$-$&$-$&$-$&$-$&$-$&$-$&$-$ \\
0219+443 &0.852&  0.89&  1.17&$W$       &14.88$\pm$0.02&   2.55&96.7&42.3&Yes&0.51&61&13&4.8&15.8$\pm$0.1&15.4$\pm$0.1&1.73$\pm$0.15\\
0226+467 &1.216&  2.45&  0.93&$O_1$&16.27$\pm$0.08 &   8.23&99.8&44.7&Yes&0.37&84& 9&9.1&17.2$\pm$0.2&16.9$\pm$0.1&5.25$\pm$0.02\\
0232+411b&0.500&  2.75&  0.99&$W$       &14.27$\pm$0.02&   7.63&99.6&34.2&Yes&0.36&84& 9&9.1&15.4$\pm$0.1&14.7$\pm$0.1&1.44$\pm$0.01\\
0249+383 &1.120&  0.95&  0.34&$W$       &16.17$\pm$0.06&   4.38&98.5&46.0&Yes&0.60&40&17&2.3&16.8$\pm$0.1&17.0$\pm$0.1&2.01$\pm$0.13\\
0255+460 &1.210&  1.74&  0.77&$O_1$&17.89$\pm$0.20&   0.30&90.2&$<$1.0&No&$-$&$-$&$-$&$-$&$-$&$-$&$-$ \\
0704+384 &0.579&  2.87&  1.06&$O_1$&15.02$\pm$0.16&  17.62&99.9&33.5&Yes&0.37&84& 9&9.1&16.2$\pm$0.1&15.5$\pm$0.2&5.91$\pm$0.02\\
0729+391 &0.663&  0.26&  0.43&$O_1$&15.60$\pm$0.14&  20.38&99.9&91.7&Yes&0.57&48&15&3.2&15.7$\pm$0.1&18.3$\pm$0.2&3.77$\pm$0.02\\
0740+380C&1.063&  5.55&  1.29&$O_2$&15.62$\pm$0.05&  23.02&99.9&35.5&Yes&0.60&40&17&2.3&16.8$\pm$0.3&16.1$\pm$0.2&4.08$\pm$0.08\\
0926+388 &1.630&  0.50&  1.09&$O_2$&17.09$\pm$0.08&   9.95&99.9&23.6&Yes&0.56&48&15&3.2&18.7$\pm$0.4&17.4$\pm$0.1&4.05$\pm$0.14\\
1123+395 &1.470&  0.36&  0.82&$O_2$&15.72$\pm$0.05&   5.45&99.1& 2.2&No&$-$&$-$&$-$&$-$&$-$&$-$&$-$ \\
1148+387 &1.303&  1.83&  0.94&$O_2$&14.93$\pm$0.05&   3.85&97.9&$<$1.0&No&$-$&$-$&$-$&$-$&$-$&$-$&$-$ \\
1206+439B&1.400&  5.69&  0.85&$O_2$&15.10$\pm$0.05&   4.85&98.8&$<$1.0&No&$-$&$-$&$-$&$-$&$-$&$-$&$-$ \\
1315+396 &1.560&  1.16&  0.49&$O_2$&15.93$\pm$0.06&   4.85&98.8&28.3&Yes&0.49&61&13&4.8&17.3$\pm$0.3&16.3$\pm$0.1&3.20$\pm$0.11\\
1435+383 &1.600&  0.54&  0.85&$O_2$&16.32$\pm$0.06&   4.25&98.3&$<$1.0&No&$-$&$-$&$-$&$-$&$-$&$-$&$-$ \\
2311+469 &0.745&  4.34&  0.69&$O_1$&14.79$\pm$0.06&   1.01&92.2& 3.5&No&$-$&$-$&$-$&$-$&$-$&$-$&$-$ \\
2332+388 &3.200&  0.42&  0.94&$O_1$&17.66$\pm$0.14&   0.68&91.0&12.5&No&$-$&$-$&$-$&$-$&$-$&$-$&$-$ \\
2344+429 &1.556&  0.92&  0.58&$O_1$&16.30$\pm$0.07&   1.10&92.7&$<$1.0&No&$-$&$-$&$-$&$-$&$-$&$-$&$-$ \\
2349+410 &2.046&  1.38&  0.98&$O_1$&16.78$\pm$0.18&   3.18&97.5&13.8&Yes&0.48&61&13&4.8&19.1$\pm$0.4&17.1$\pm$0.2&3.35$\pm$0.02\\
2351+456 &2.000&  2.21&  0.27&$O_1$&16.03$\pm$0.16&   6.39&99.4&17.9&Yes&0.39&84&11&7.7&18.0$\pm$0.2&16.4$\pm$0.2&2.54$\pm$0.08\\
\hline
\end{tabular}
\end{center}

(1) Object name in the B3-VLA sample. (2) Object redshift (Vigotti et al. 1997). 
(3) Radio flux in Jy at 408 MHz. (4) Radio spectral index, $\alpha_{\rm 408}^{\rm 1460}$, 
defined as $S_{1460}=S_{408}\nu^{-\alpha_{\rm 408}^{\rm 1460}}$. (5) Observing run: $W$, September 1996, WHT La Palma. 
$O_1$, October 1997, 3.5m Calar Alto. $O_2$, February 1999, 3.5m Calar Alto. 
(6) $K$-band aperture photometry. (7) $\chi^2/\nu$ of the surface brightness fitting to 
the PSF model. (8) percentage probability of being extended, derived from the previous 
$\chi^2/\nu$. 
(9) Contribution of the possible extension to the total flux.
(10) Classification output (Yes=Extended, No=Not Extended). 
(11) Free parameter of the generalized model ($\alpha$=1 for disk galaxies, $\alpha$=0.25 for elliptical galaxies).
(12) percentage probability of being an elliptical galaxy. (13) percentage probability of being a disk galaxy.
(14) Ratio of both probabilities. (15) Host $K$-band magnitude. (16) Nucleus $K$-band magnitude.
(17) Effective radius of the Host in arcsec.

\end{sidewaystable*}

We have assumed that any extended emission detected around the quasars
is associated with the stellar component of the HG. Other types of
extended emission, such as radio-aligned components similar to those
seen in radio galaxies (Dunlop \& Peacock 1993) or synchrotron-beamed
emission, might be detected in the $K$-band, although they are
expected to contribute less than 10\% to the HG flux (Riegler et al.
1992; Ridgway \& Stockton 1997). For the objects at $z>$2 some
contribution from H$_{\alpha}$ is expected, about $\sim$15\% (Lehnert
et al. 1999b). Therefore, we assume that the images of the quasars at
the NIR wavelengths consist of two components only: (i) the nuclear,
point like source, characterized by a scaled PSF (to match its flux),
and (ii) the host galaxy.

\begin{figure}
\centerline{\vbox to 0cm{\vfil}
\includegraphics[angle=270,width=9cm]{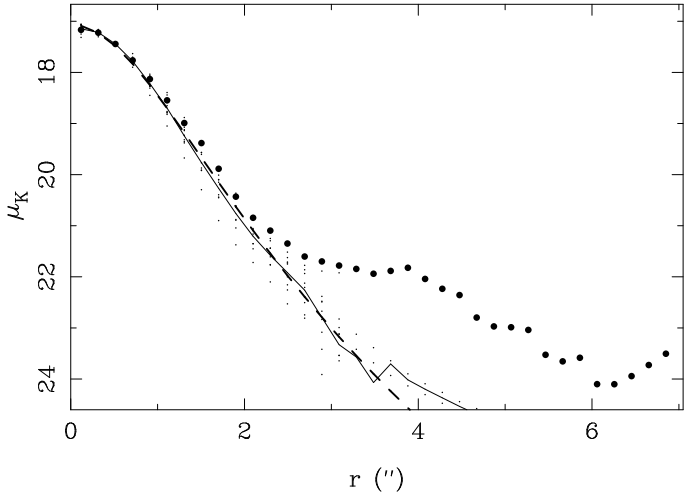}
}
\caption[Surface brightness profile (SBP) of the quasar B3 0740+380C (solid circles) together
with the SBP of the field stars scaled to the quasar intensity peak (points).
The solid line shows the mean of the stars' SBPs, and the dashed line shows the
mean PSF derived from fitting a Moffat function.]
{
Surface brightness profile (SBP) of the quasar B3 0740+380C (solid circles) together
with the SBP of the field stars scaled to the quasar intensity peak (points).
The solid line shows the mean of the stars' SBPs, and the dashed line shows the
mean PSF derived from fitting a Moffat function.
 }
\label{fig:example}
\end{figure}

Our procedure to detect and recover the HG consists on the following
steps: 

\begin{enumerate}
  
\item We obtained for each image the surface brightness profiles (SBP)
of the quasar and of a number of field stars. We masked the possible
contaminating sources before obtaining the SBP.  This includes
basically all the sources that are more than 3\arcsec\ distant from
the quasar peak and that do not show any connection such as bridges or
tails with the quasar.  To obtain the SBP increases the signal to
noise ratio since it implies an average of the brightness along the
eccentric anomaly. We reduce the problem to one dimension, although
this implies a loss of information. This procedure allowed us to study
the properties of HGs that were initially poorly visible in the
images. The SBP was obtained using the techniques discussed in
Jedrzejewski (1987), implemented in the ELLIPSE task included in the
stsdas package of IRAF.

\item The PSF was modelled by a Moffat function (Moffat 1969). The use
of an analytical model allows us to give a better description of the
PSF in the outer regions (normally more noisy) and let us use the PSF
for a proper convolution with galaxy models (see below).  To determine
the best PSF model for each image we fitted the SBP of the field
stars to the Moffat function

$$I_{\rm Moffat}(r;I_o,r_{\rm 1/2},\beta)=I_o \left[ 1
+(2^{1/\beta}-1)\cdot(\frac{r}{r_{\rm 1/2}})^2  \right]^{-\beta}$$

The width of the central core is determined by $r_{\rm 1/2}$, and
$\beta$ determines the extension of the wings. $I_o$ is the scaling
factor that determines the peak of the intensity. Once the field stars
are fitted to this model we obtain a mean value for both $r_{\rm 1/2}$
and $\beta$. These values characterize the {\it mean} PSF in the
image.

It was found that about $\sim$10 stars per field were enough to
determine the PSF parameters. There were enough stars per field in the
OMEGA images, due to their large field of view. However, the WHIRCAM
images have only 1 or 2 stars per field; still, these were found to
present a stable PSF image-to-image, mainly due to the short time
period in which they were obtained ($<$2.5 hours). Therefore, we used
all the stars available in all the WHIRCAM images to determine the PSF
parameters.

\item The {\it mean} PSF was used to determine the probability that a
quasar was extended by fitting the surface brightness profile of the
quasar to the function: $PSF (r;I_o)= I_{\rm Moffat}
(r;I_o,\hat{r_{\rm 1/2}},\hat{\beta})$. This method was tested on 200
simulated stars, with the same characteristics of seeing, brightness
and signal-to-noise as the observed quasars. This yields a
probability distribution that the surface brightness of a star fits 
the {\it mean} PSF, a distribution that can be used to estimate the
probability that a quasar is extended or not. We have applied the
same test to the field stars, finding that they fit  the mean PSF
model within a 95\% probability, as expected. We also found that 
when an artificial extension is detected in a field star, it
always contribute less than 10\% to the total star flux.

Figure \ref{fig:example} ilustrates this procedure. We show the SBP of
the quasar B3 0740+380C, together with the scaled SBP of the field
stars used to determine the mean PSF. There is a dispersion of the
individual star SBPs around the mean PSF. The quasar profile is further
from the mean PSF than the field stars, and shows a clear
extension. This difference in the SBP compared with the differences
found for the stars can be used to set a probability for the extension
to be real.

\item The surface brightness profile of the quasars was then fitted to
a two-component model: a free-scale {\it mean} PSF model plus a
galactic model. We have initially restricted ourselved to the two most
commonly used analytical prescriptions for the galactic model: an
exponential Freeman (1970) law describing disk galaxies and a de
Vaucouleurs (1948) $r^{1/4}$ law describing elliptical galaxies.  The
fitting procedure has three free parameters: the nuclear component
flux, the host galaxy flux, and the galaxy scale (for each analytical
model).  The galactic profile was convolved during each step of the
fitting procedure by the mean PSF (normalized to one), in order to
take into account the effects of the seeing on the shape of the
profiles. In general, the nuclear component has a more intense surface
brightness, with larger signal-to-noise ratio. Therefore, we assume
that with this method we are able to recover the flux of the nuclear
component better than that of the extended component. We then estimate
the HG flux by subtracting the recovered nuclear component flux from
the total flux.

\item We found that the quasar profile fits both models well.  In
  order to investigate this we applied the procedure to a sample of
  1000 simulated quasars. These simulated quasars consist on a central
  point-like source, following a Moffat function, plus an extended source,
  following an exponential or a $r^{1/4}$ function.  These images cover
  different ranges of the paramter space, such as different
  ratios between the central and the extended emission, between the effective
  radius and the seeing, and the signal to noise of the total source.  They
  reproduce the basic parameters of the real data, such as  background noise,
  pixel scale, and image depth. The simulations are explained in S\'anchez
  (2001) and they will be published in a separate paper (S\'anchez 2003).
 
 They show that it is not possible in our range of parameters to
 distinguish between a spiral or an elliptical HG by the comparison of
 the $\chi^2/\nu$ resulting from the profile fitting to the two
 above-mentioned models. This problem has been described before
 (McLure et al. 1999). This could be due to the larger signal-to-noise
 of the central region, which dominates the fits to such an extent
 that subtle differences in the extended emission become
 indistinguishable. We also found that the percentages of flux
 recovered for both the HG and the nuclear source are quite similar,
 independent of the model.

\item To solve this problem we followed a different approach: (i) the
HG surface brightness profile was obtained subtracting the {\it mean}
PSF model, scaled to the nuclear source flux (previously obtained),
and (ii) this profile was fitted to a generalized galaxy model
($I_{\rm gal}\propto e^{r^{\alpha}}$). This model includes the other
two as particular cases: the exponential model is recovered if
$\alpha$=1, and the $r^{1/4}$-model if $\alpha$=1/4. The values of
$\alpha$ obtained with this method were used to classify the HGs.
This method was applied to the simulated images, and yielded sharply
different distributions of the $\alpha$ parameter for the spiral and
for the elliptical galaxies (Figure \ref{fig:prob_alfs}).  We have
used the obtained $\alpha$ to classify the HG morphologically.  A
similar procedure has been applied by McLure et al. (1999) to
distingue between exponential and $r^{1/4}$ models.

The idea behind this method is that the most important region to
  distinguish between an elliptical and a disk galaxy is the galaxy
  core.  The core is severely affected by the nuclear component
  substraction. Even in the case of a perfect subtraction there is an
  increase of the noise: The photon noise in this area is the sum of
  the photon noise from the galaxy and the QSO nucleus. Therefore,
  there is a loss of information in the inner part.  Indeed, it is
  noticed in the simulations that the preferred $\alpha$ parameters of
  a simulated elliptical or spiral galaxy were not the expected values
  (0.25 and 1.0, respectively) but somewhat larger ones (0.3 and 1.15,
  respectively). Effects like this have been noticed by other authors
  (Meurer et al.  1995; Whitmore et al. 1999).  If we force the model
  to fit the data to the theoretical values, both of them will fit
  equally badly, and cannot distinguish between the models. However,
  letting $\alpha$ free to be fitted we can recover the parameter that
  fits the data best.

\end{enumerate}

\begin{figure}
\centerline{\vbox to 0cm{\vfil}
\epsfysize=9cm
%\epsffile{prob_alf.ps}}
\epsffile{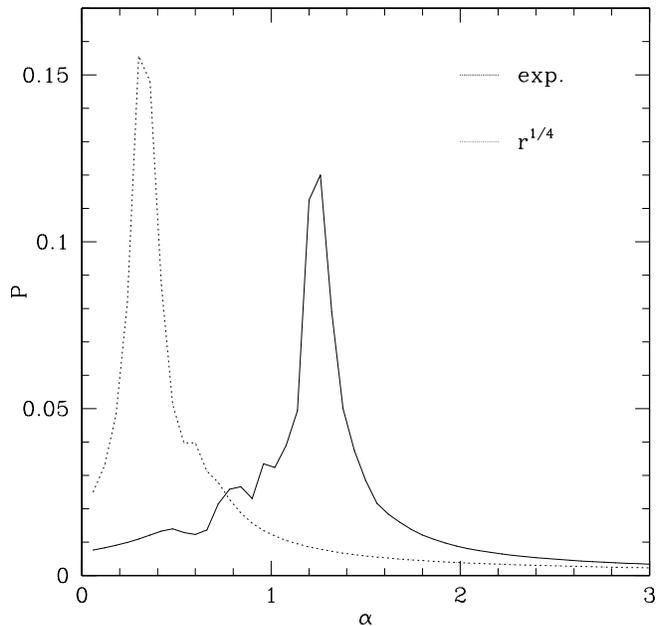}}
\caption[Probability density functions for exponent $\alpha$ 
obtained from fits to simulated elliptical and disk galaxies.]
{Probability density functions for exponent $\alpha$ obtained from fits 
to simulated elliptical and disk galaxies.}
\label{fig:prob_alfs}
\end{figure}

In Paper I we used a rather different method to detect and restore the
HGs. Before the surface brightness profile of the quasars was
obtained, the images were deconvolved by a mean PSF. The brightness
profiles were then obtained and fitted to different models: a gaussian
function (which characterizes the deconvolved point-like sources) and
a gaussian plus galactic model (which characterizes a point-like plus
an extended source). This method had to be applied due to the lower quality
of the data used in Paper I. In order to compare the results obtanied
in Paper I with the present results, and to set limits to both methods, we
have applied both of them to the simulations mentioned above.

We found that both methods are able to recover the flux of the HGs,
although in different ranges of parameters. The method used in this
article is valid for HGs for which the contribution to the total flux
was larger than $\sim$9\% ($\sim$12\% for the method used in Paper
I). In this range of fluxes, the magnitudes of the HGs are recovered
with an error of about $\sim$0.26 mag ($\sim$0.52 mag for the method
used in Paper I). There is no tendency to over- or underestimate the
flux down to $K_{\rm HG}\sim$18.5 mag ($\sim$17 for Paper I).
Moreover, with the method presented here it is possible to recover the
effective radius of the HG, with an error of $\sim$25\% (A parameter
that we could not recover in Paper I). However, the current method
requires (i) a good sampling of the PSF, using $\sim$5-7 stars with
the same brightness as the quasars, or a more reduced number of
brighter stars, and (ii) that the sizes of the HGs are similar to or
larger than the seeing, which, for quasars at $z\sim$1 limits the use
of this method to good-seeing images (i.e., better than
$\sim$1.3$\arcsec$). The method used in Paper I, based on a
deconvolution process, is still valid for lesser-quality images. It
does not require a precise shampling of PSF and it can be used for
images with seeing larger than 1.3$\arcsec$, although it recovers the
flux of the HG with larger uncertainties (0.52 mag) and for brighter
HGs ($>$12\%).

\section{Results of the analysis}

Table \ref{tab:Sample} lists the sample of objects, including some
basic parameters, such as the redshift and radio properties, and the
$K$-band photometry. We have also listed the results from the analysis.
A QSO has been classified as extended whenever (1) the probability that
its SBP deviates from the mean PSF profile is larger than 95\% 
and (2) the contribution of the extended source to the total flux
is larger than 9\%. These criteria have been selected based on the
results from the simulations and our analysis over the field stars.

We found an extension in 16 of the 31 quasars (55\% of the
sample). The extension around B3 0143+446 is considered dubious,
although it satisfies the criteria. This object shows a nearby star at
less than 6$\arcsec$ which could affect its SBP even when a proper
mask has been applied. We will not include this object in our further
analysis.  Ten QSOs have a probability of being extended of more than
99\%.  The remaining 6 quasars have probabilities between 95\% and
99\%. Of theses 6, four are in the range of the faintest detected HGs
($K_{\rm HG}>18$ mag). The fraction of detected HGs decreases with
redshift, being 75\% of the sample at $z<$1 and 48\% at $z>$1. Beyond
this redshift the fraction of detected galaxies is roughly
constant. This result is expected since at high redshift both the
surface brightness and apparent luminosity of the HGs decrease, which
makes their detection difficult. We have not set a magnitude limit for
the undetected host galaxies, since the detection does not depend only
on the brightness of the source.

All the HGs have a larger probability of being elliptical galaxies
than of being spirals. For an individual HG the probability that it is
a spiral is a 7-17\% and that it is and elliptical is a 40-84\%.  This
is the first time that a morphological analysis of the HGs of a sample
of radio quasars yields a conclusive result beyond $z>$0.4.  Previous
results (Kotilainen \& Falomo 2000, Kotilainen et al. 1998, Kukula et
al. 2001) directly assume a certain profile ($r^{1/4}$), or have
inconclusive results from their analysis. Our result shows that the
HGs of radio quasars at high redshift have the same morphology as the
radio galaxies, supporting the unification schemes of both radio
sources.  Similar results were found by Taylor et al.  (1996), McLure
et al. (1999) and Dunlop et al. (2001), at low redshift ($z<$0.4).

The photometry of seven of these objects was presented in Paper I. The
differences between the $K$-band magnitudes are within the errors
($\sim$0.3 mag). In four of them it was possible to follow a
morphological analysis. Three have an extension detected in both
studies (B3 0704+384, B3 0740+380C and B3 1315+396). The differences
between their HG magnitudes are -0.5, -0.4 and 0.7 (respectively),
with a mean difference of 0.05$\pm$0.65 mag. These values are within
the expected errors ($\sim$0.5 mag for Paper I HG magnitudes, and
$\sim$0.2 mag for current data). The extension of the remaining quasar
(B3 0926+388) has been detected in the present study only, since its
magnitude is much fainter than the detection limit for HGs reported in
Paper I.

The HG contribution to the total flux ranges between 14\% (B3
2349+410) and 92\% (B3 0729+391), with a mean contribution of
$\sim$41\%. Similar values have been reported in Paper I (range
between 18\%-83\% and mean $\sim$50\%), and studies based on lower
redshift samples for similar wavelengths (Taylor et al. 1996,
Kotilainen et al. 1998). These values are lower at optical
wavelengths, $\le$20\% for $R$-band images (e.g., Lehnert et al.
1999a, McLure et al. 1999). Therefore, the extended emission is redder
than the nuclear emission. This result is expected if a stellar
component dominates the extended emission rather than other
contributions (such as scattered flux from the nucleus). Recent
results based on off-nuclear spectroscopy have confirmed its stellar
nature (Canalizo \& Stockton 2000; Nolan et al.  2001; Canalizo \&
Stockton 2001; Courbin et al.  2002).

We have obtained a restored image of the HG by subtracting an image of
the nuclear point-like source. This image was built assuming a Moffat
function with the mean-PSF parameters (obtained by fitting to the
field stars). Its flux was fixed to the value obtained from the
fitting procedure for the nuclear component. The image was built using
the package ARTDATA implemented in IRAF. Figure \ref{fig:ima_1} shows
the contour plots and surface brightness profiles for the sources with
detected extension. In each panel (a) shows the contour plot (left)
and profile (right) of the original source, and (b) shows the contour
plot (left) and profile (right) of the recovered host galaxy. The
solid line in (a) shows the surface brighteness profile of the PSF
scaled to the quasar peak. The $\chi^2/\nu$ from the comparison of
both profiles and the probability of being extended have also been
included for each quasar. It is clearly seen that the objects present
a significant extension from the comparison of both profiles. The
orientation of the contour plots is North (up) and East (left), and
the field-of-view corresponds to 16$\arcsec$$\times$16$\arcsec$. We
have plotted the raw data before masking for nearby companions. In
some cases (e.g., B3 0006+397, B3 0110+401) there are close companions
that affect the unmasked SBP.  However, the extension is clearly seen
in spatial regions not affected by these companions.  This is not the
case for quasar B3 0143+466, classified as dubious.

\begin{figure*}
\centering
\begin{minipage}{17cm}
\begin{center}
\epsfxsize=8cm
\begin{minipage}{\epsfxsize}{\epsffile{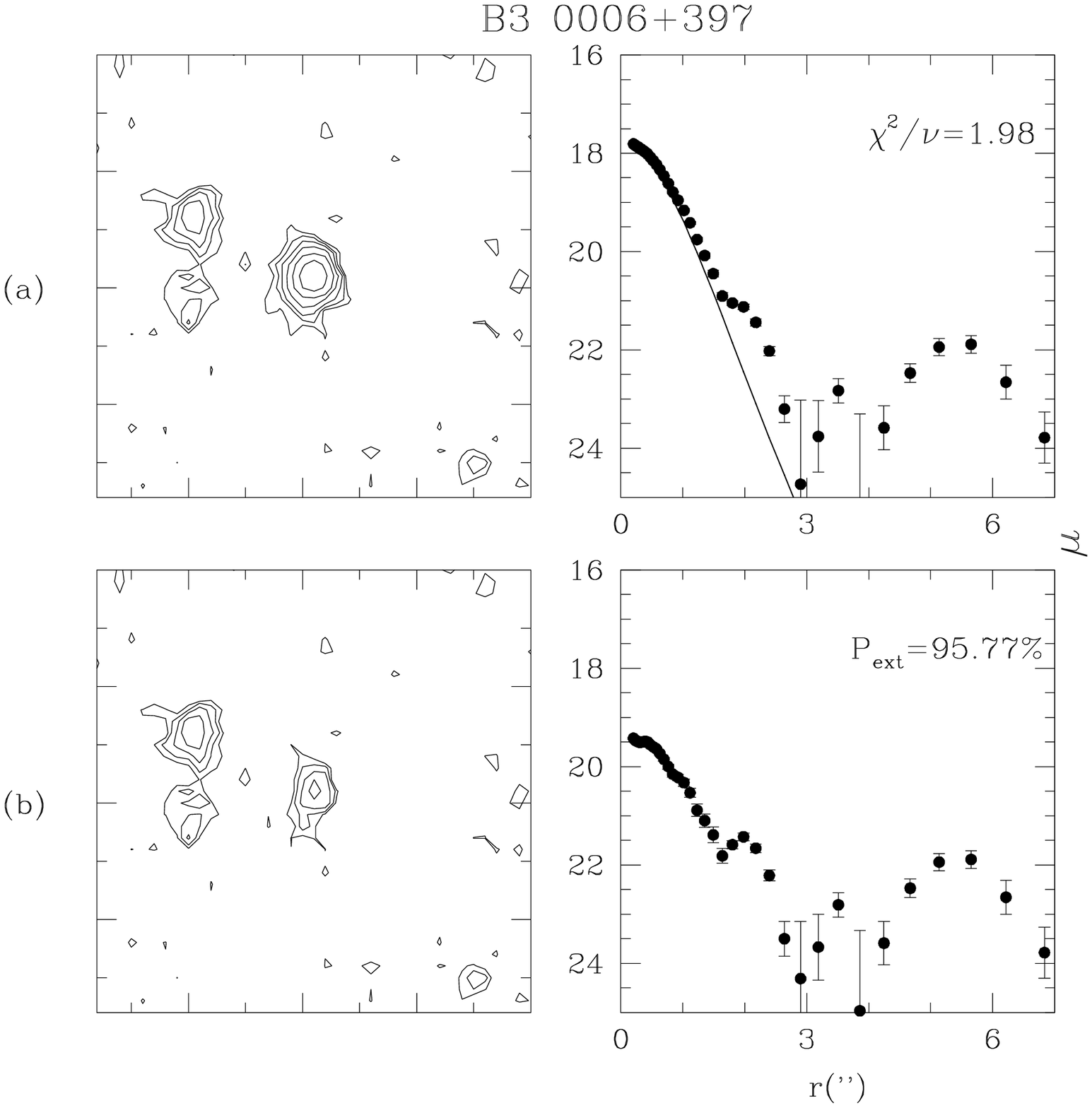}}
\end{minipage}
\epsfxsize=8cm
\begin{minipage}{\epsfxsize}{\epsffile{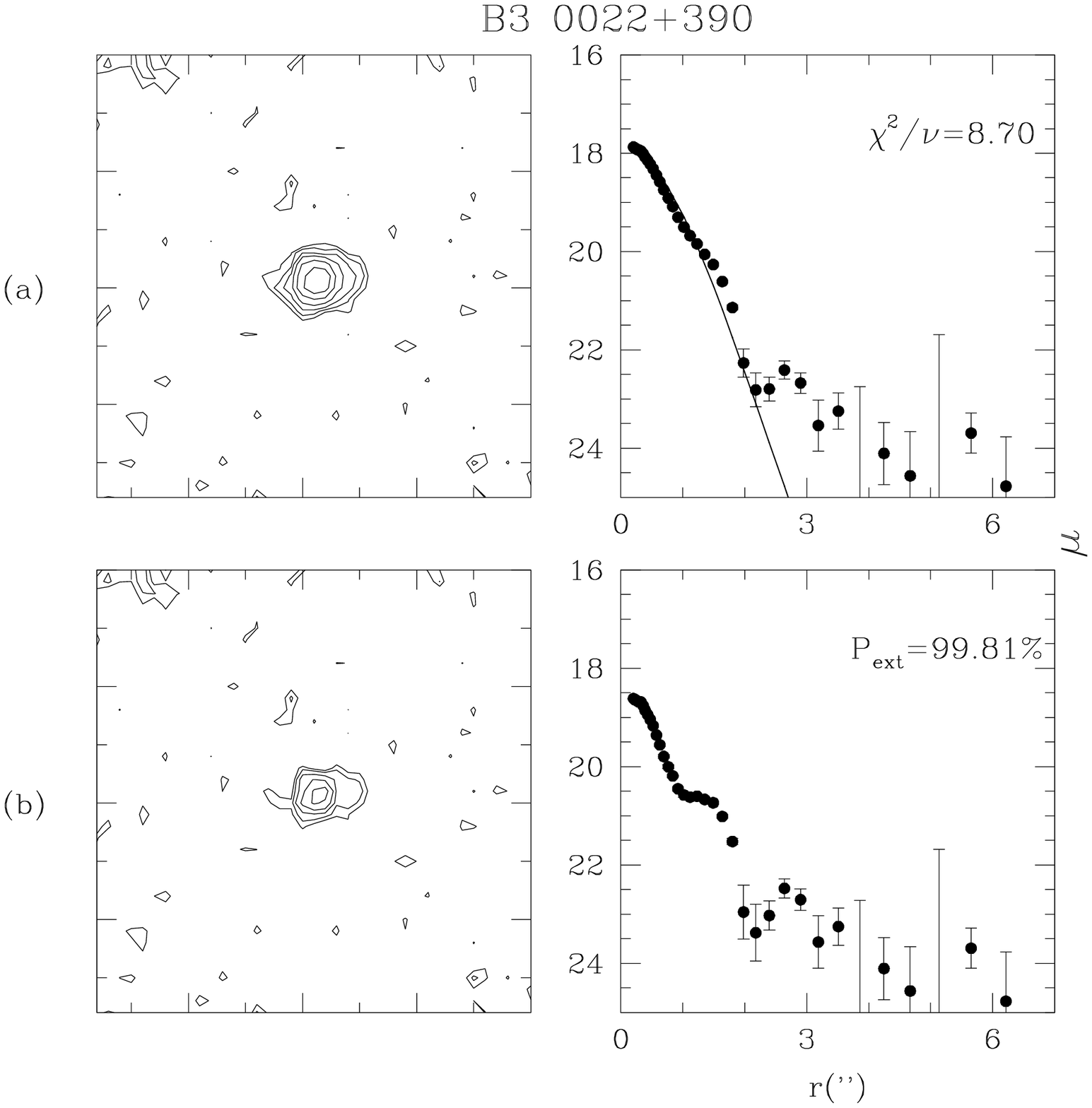}}
\end{minipage}
\epsfxsize=8cm
\begin{minipage}{\epsfxsize}{\epsffile{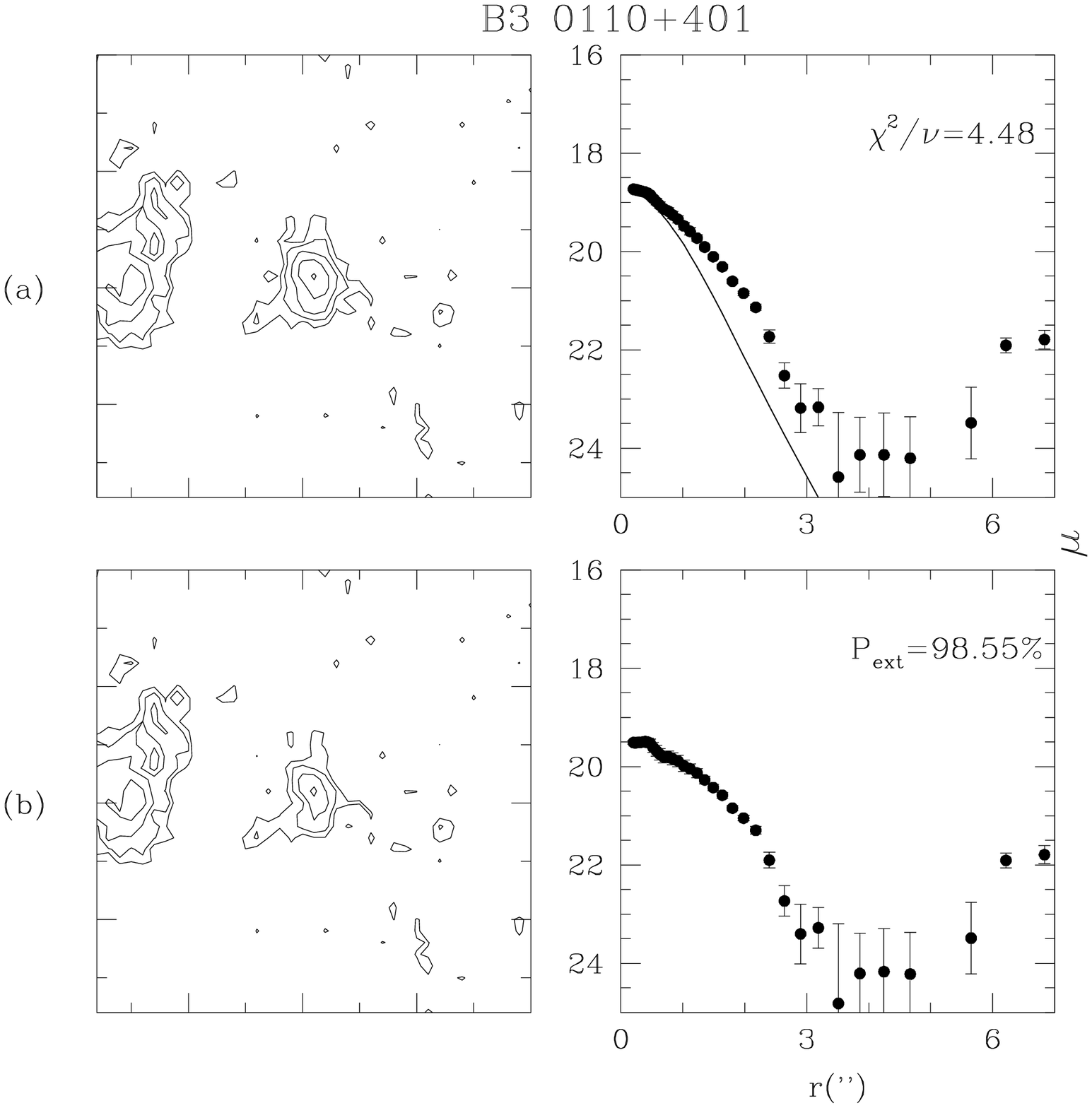}}
\end{minipage}
\epsfxsize=8cm
\begin{minipage}{\epsfxsize}{\epsffile{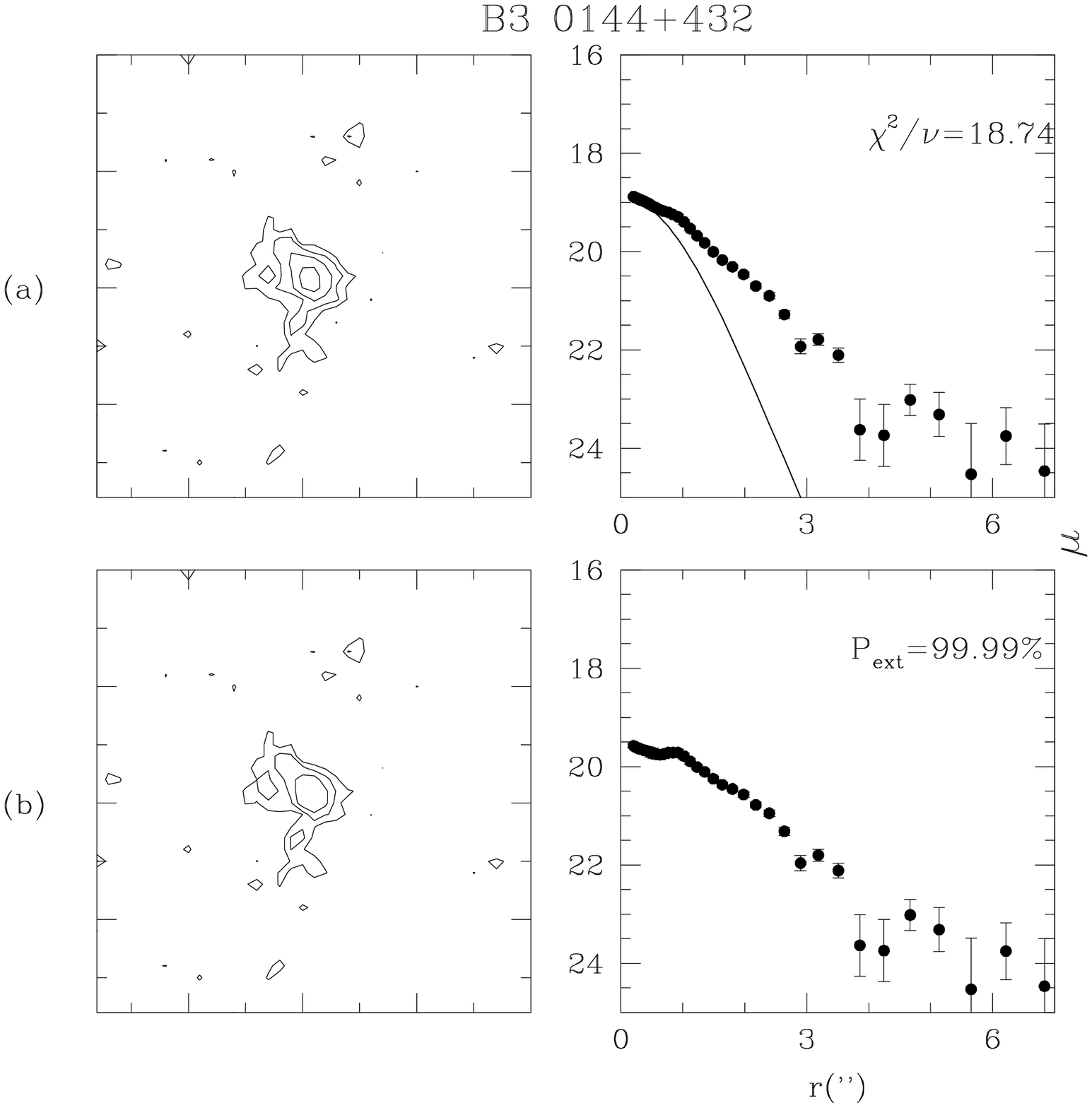}}
\end{minipage}
\end{center}
\caption[]{{\small 
Contour maps and surface brightness profiles of the quasars with detected 
extension. In each panel (a) is the contour map (left) and surface
brightness (right) of the original image, and (b) is the contour map
(left) and surface brigthness (right) of the HG image. The solid line
in (a) shows the mean PSF scaled to the quasar intensity peak. 
The distances between small marks are 2$\arcsec$ (for objects
observed with OMEGA), and 1.2$\arcsec$ (for objects observed with
WHIRCAM). The size of the contour maps is 16$\arcsec$$\times$16$\arcsec$ for all the
images. First contour level is at 2$\sigma$ per pixel, with a
separation of 0.5 mag between successive contours.
}}
\label{fig:ima_1}
\end{minipage}
\end{figure*}

\addtocounter{figure}{-1}

\begin{figure*}
\centering
\begin{minipage}{17cm}
\begin{center}
\epsfxsize=8cm
\begin{minipage}{\epsfxsize}{\epsffile{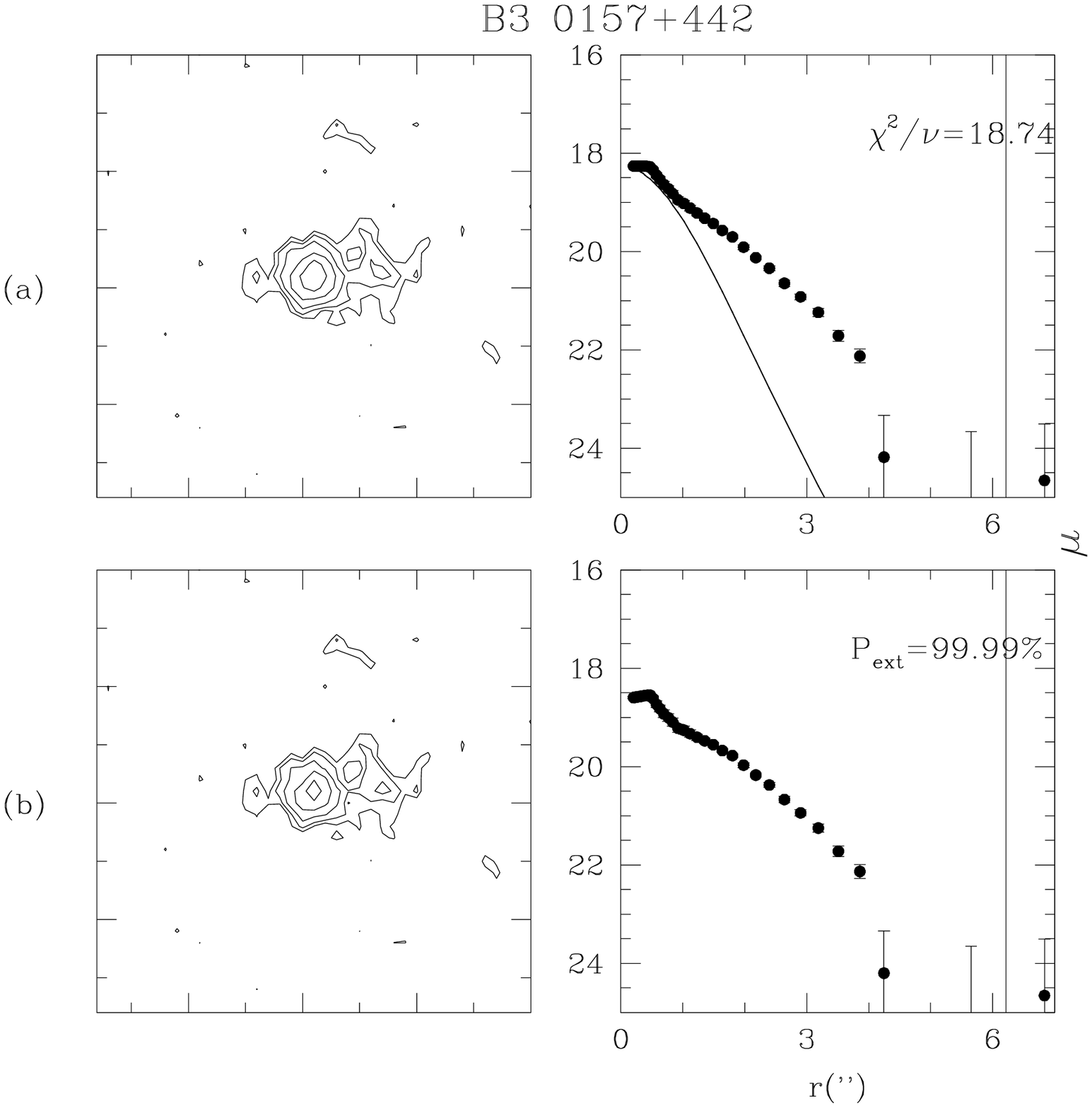}}
\end{minipage}
\epsfxsize=8cm
\begin{minipage}{\epsfxsize}{\epsffile{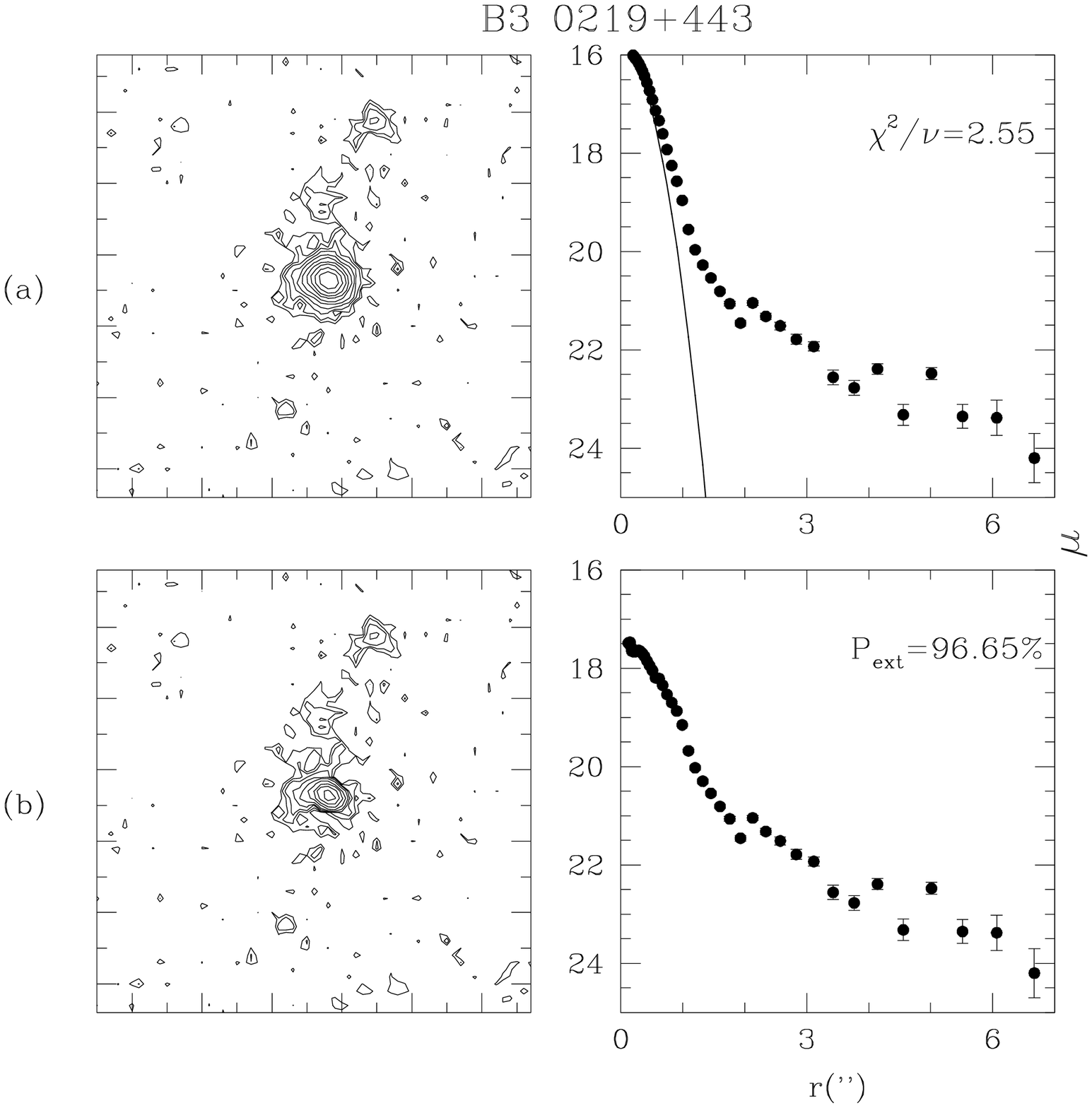}}
\end{minipage}
\epsfxsize=8cm
\begin{minipage}{\epsfxsize}{\epsffile{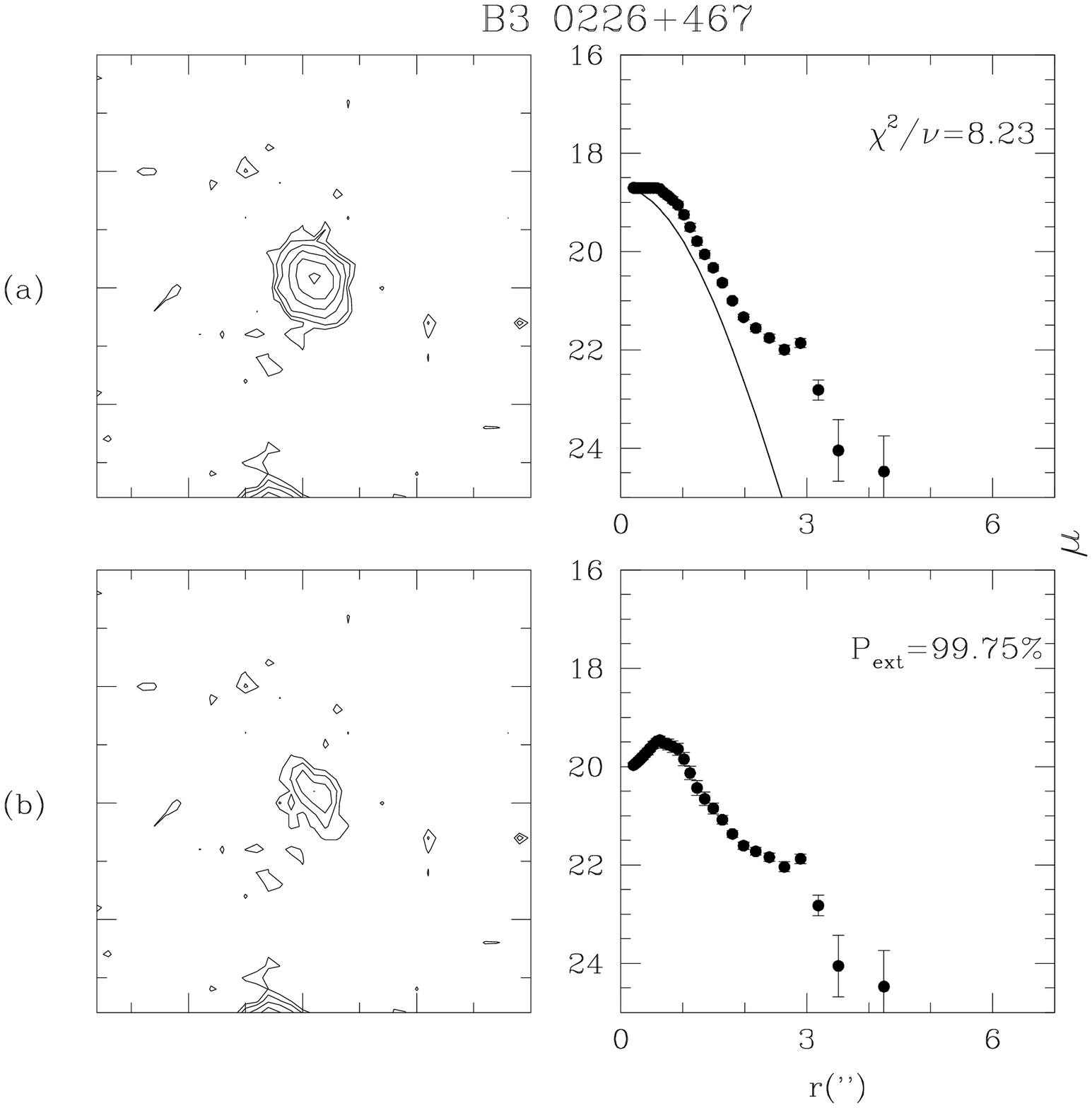}}
\end{minipage}
\epsfxsize=8cm
\begin{minipage}{\epsfxsize}{\epsffile{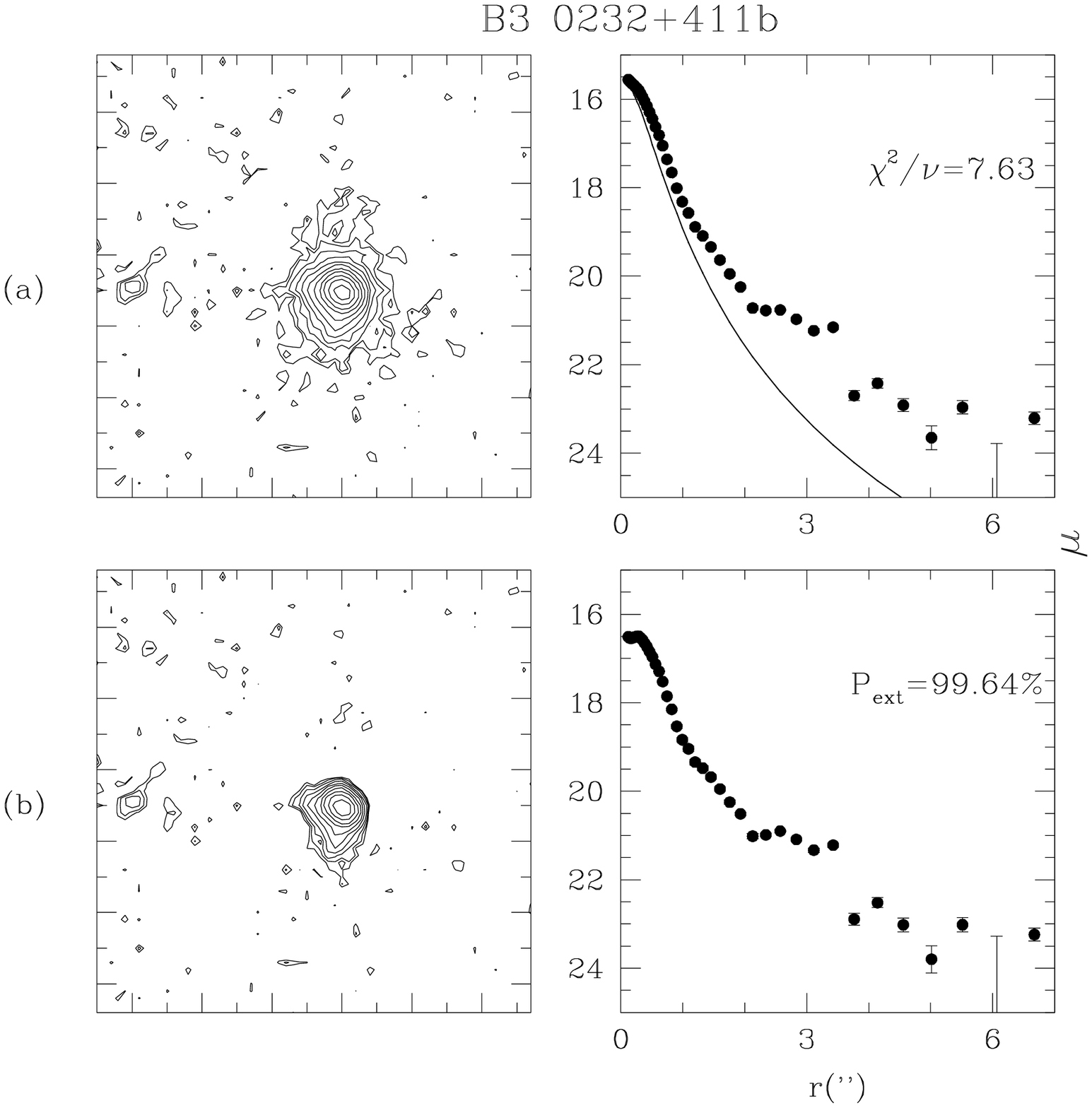}}
\end{minipage}
\end{center}
\caption[]{{\small Continued}}
\label{fig:ima_2}
\end{minipage}
\end{figure*}

\addtocounter{figure}{-1}

\begin{figure*}
\centering
\begin{minipage}{17cm}
\begin{center}
\epsfxsize=8cm
\begin{minipage}{\epsfxsize}{\epsffile{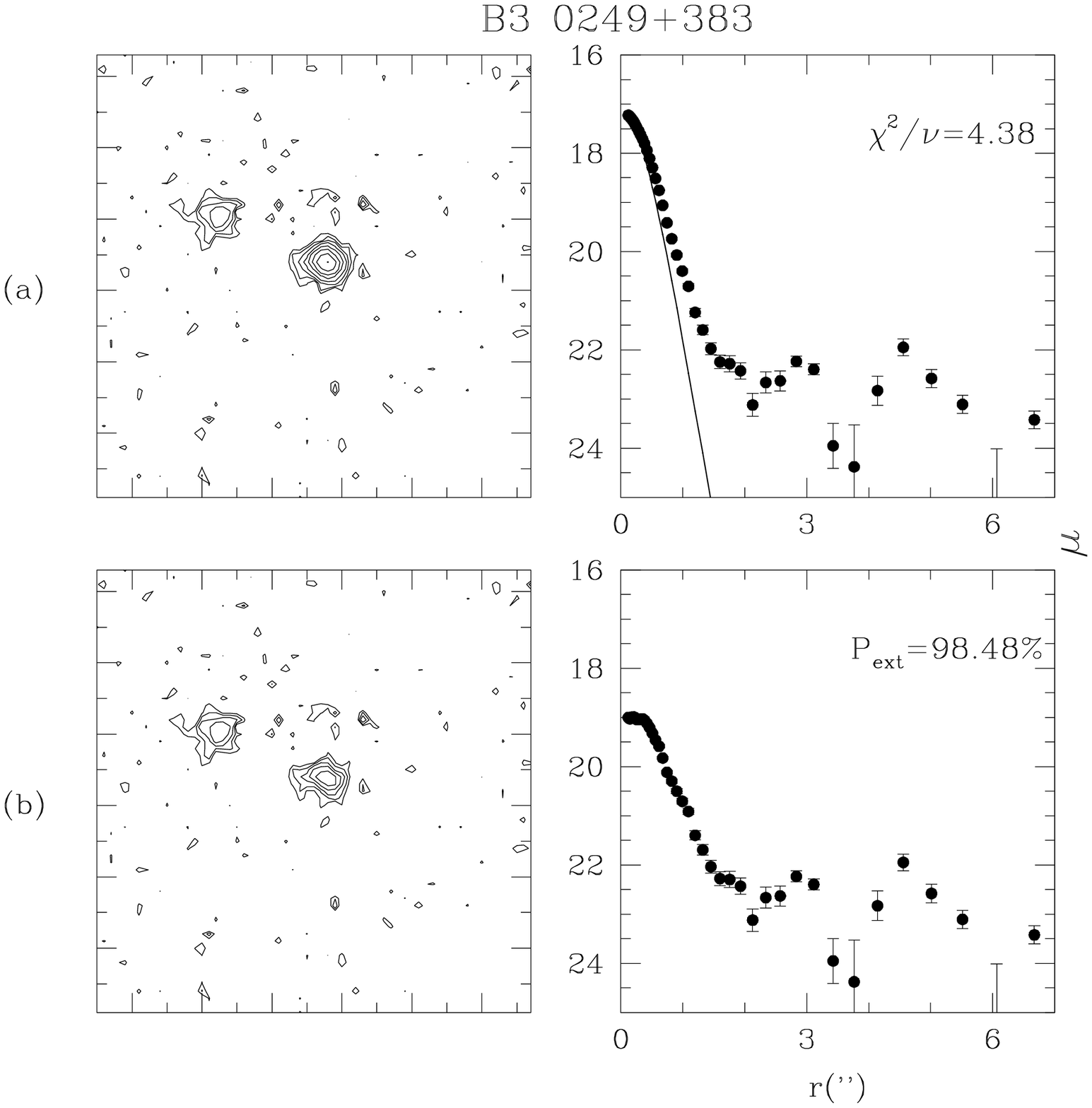}}
\end{minipage}
\epsfxsize=8cm
\begin{minipage}{\epsfxsize}{\epsffile{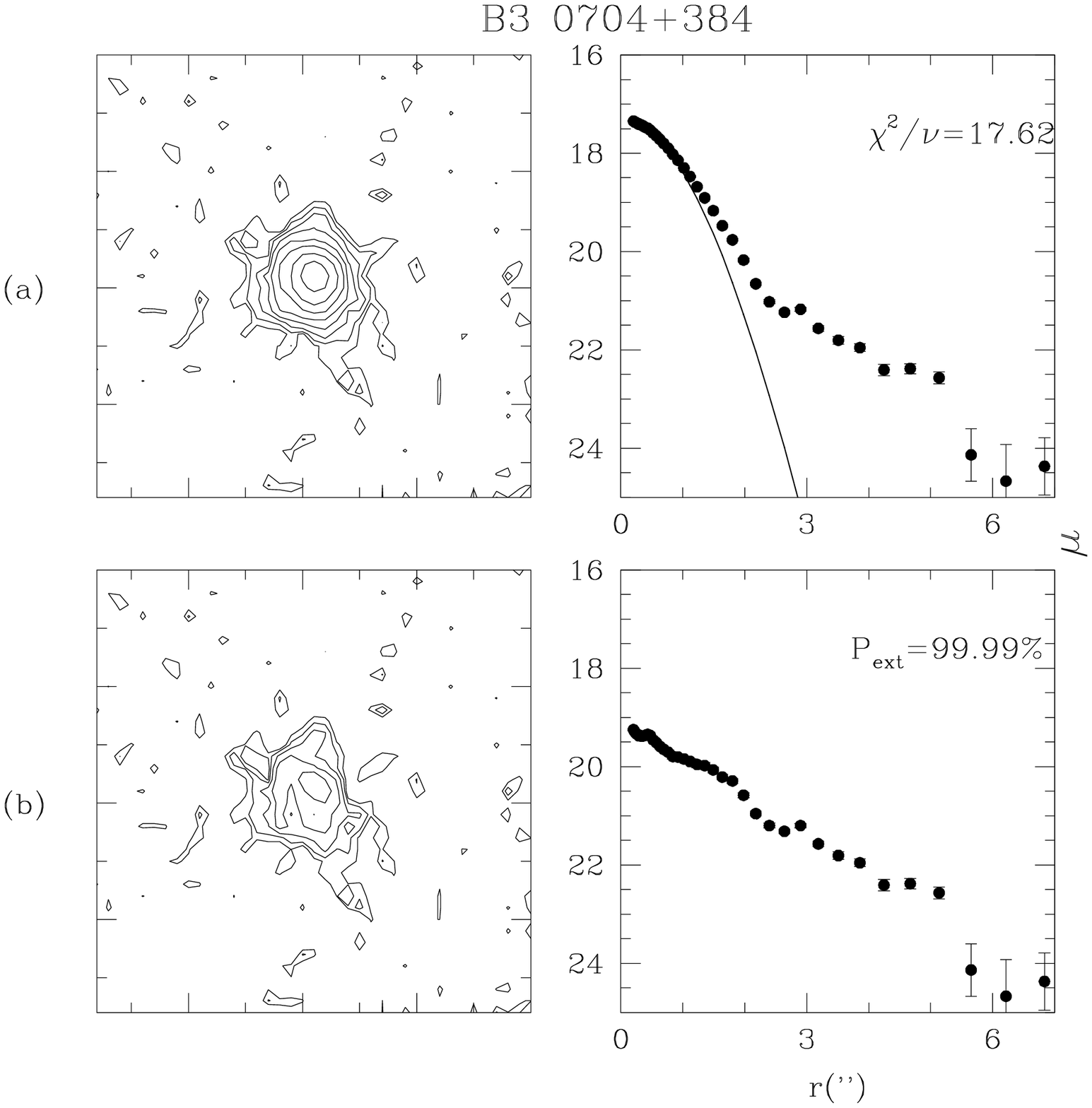}}
\end{minipage}
\epsfxsize=8cm
\begin{minipage}{\epsfxsize}{\epsffile{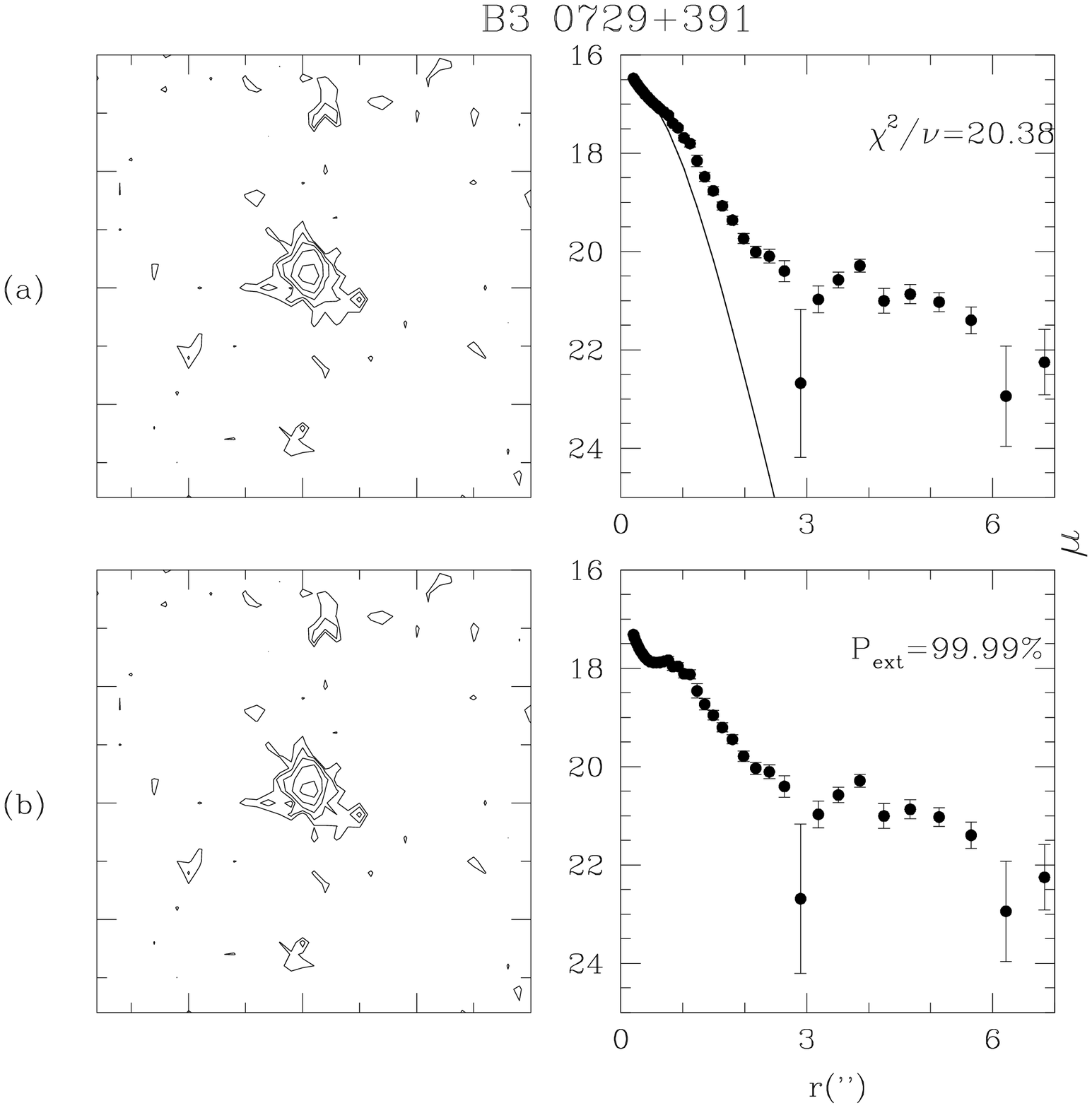}}
\end{minipage}
\epsfxsize=8cm
\begin{minipage}{\epsfxsize}{\epsffile{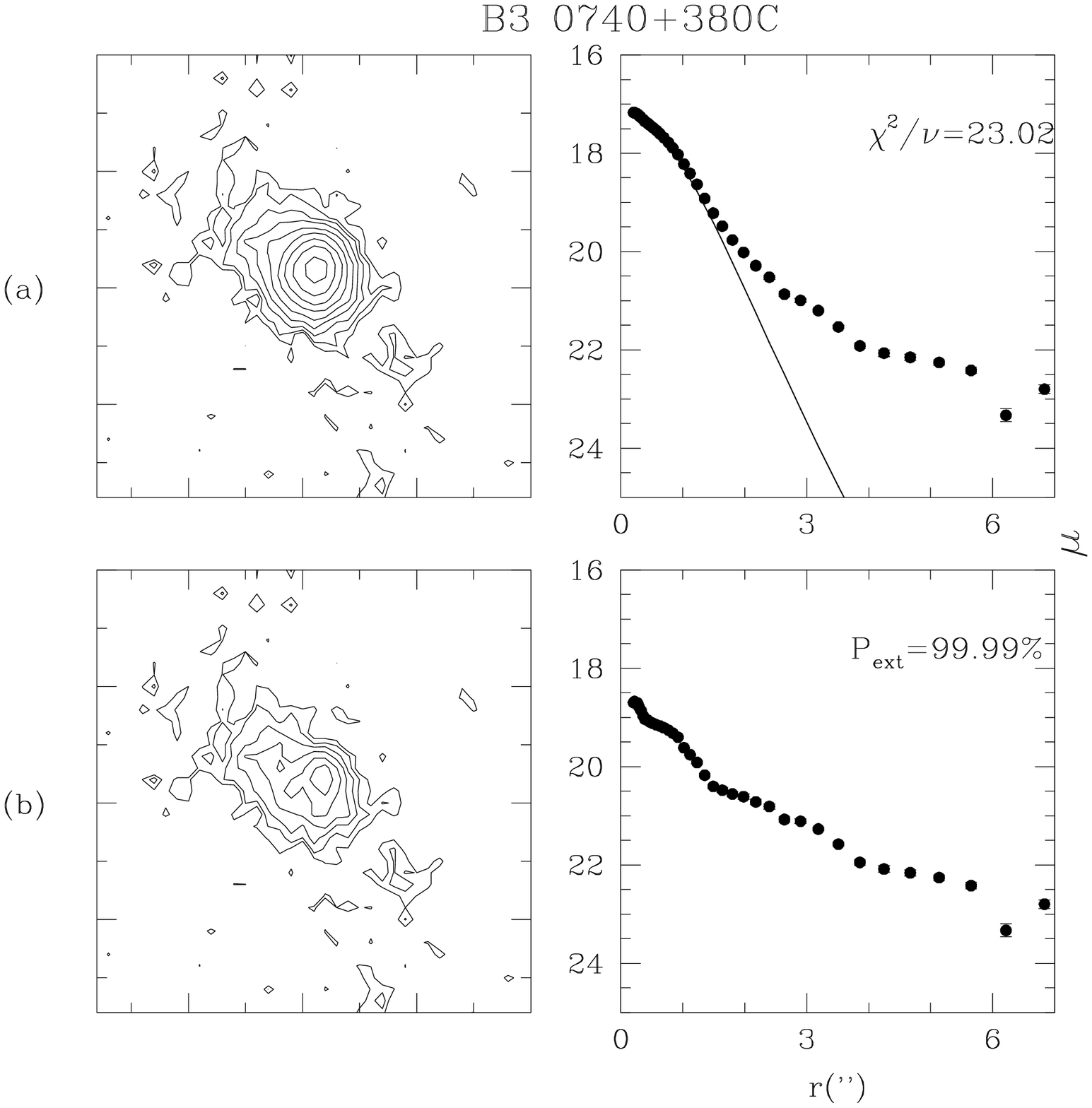}}
\end{minipage}
\end{center}
\caption[]{{\small Continued}}
\label{fig:ima_3}
\end{minipage}
\end{figure*}

\addtocounter{figure}{-1}

\begin{figure*}
\centering
\begin{minipage}{17cm}
\begin{center}
\epsfxsize=8cm
\begin{minipage}{\epsfxsize}{\epsffile{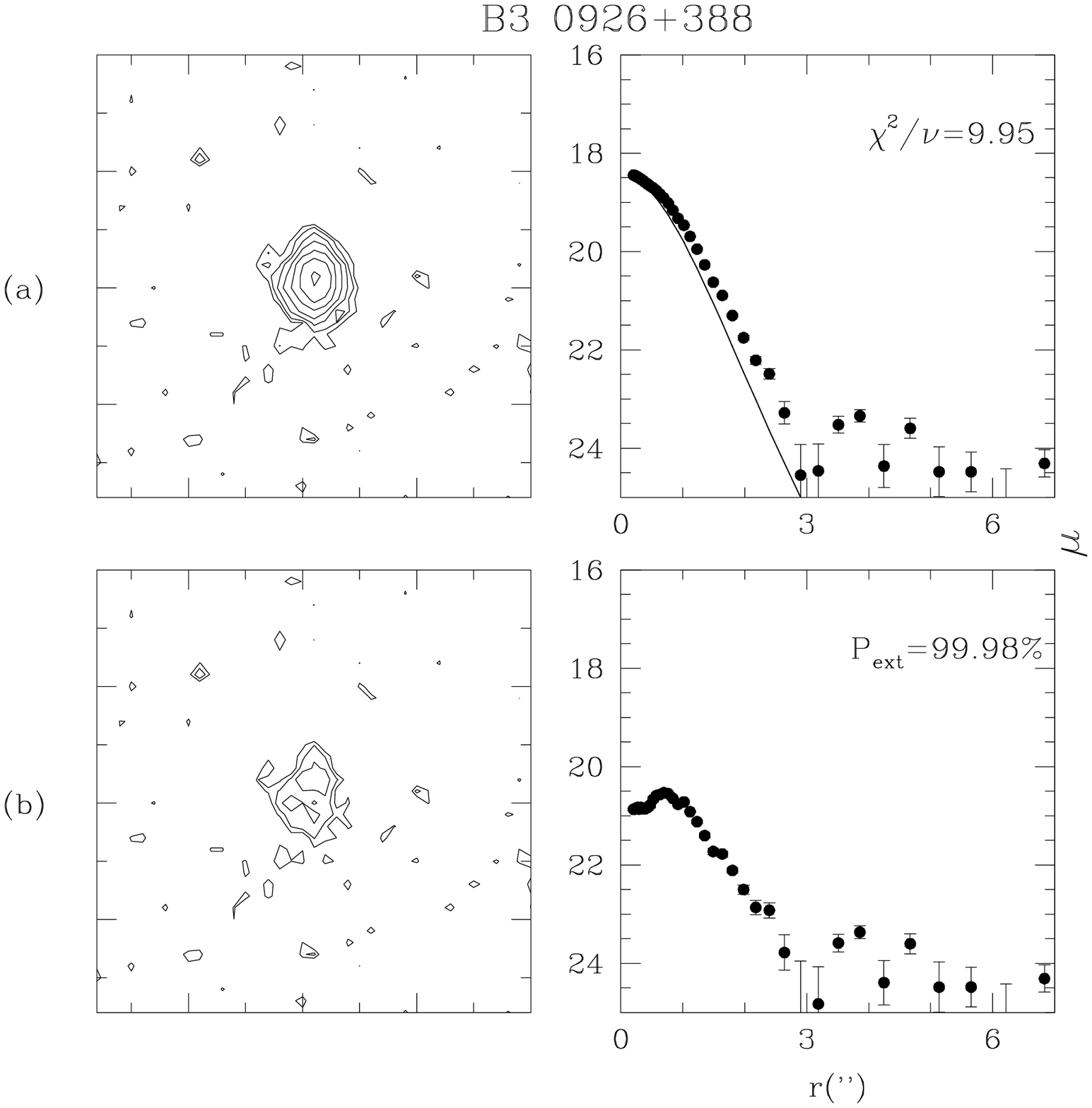}}
\end{minipage}
\epsfxsize=8cm
\begin{minipage}{\epsfxsize}{\epsffile{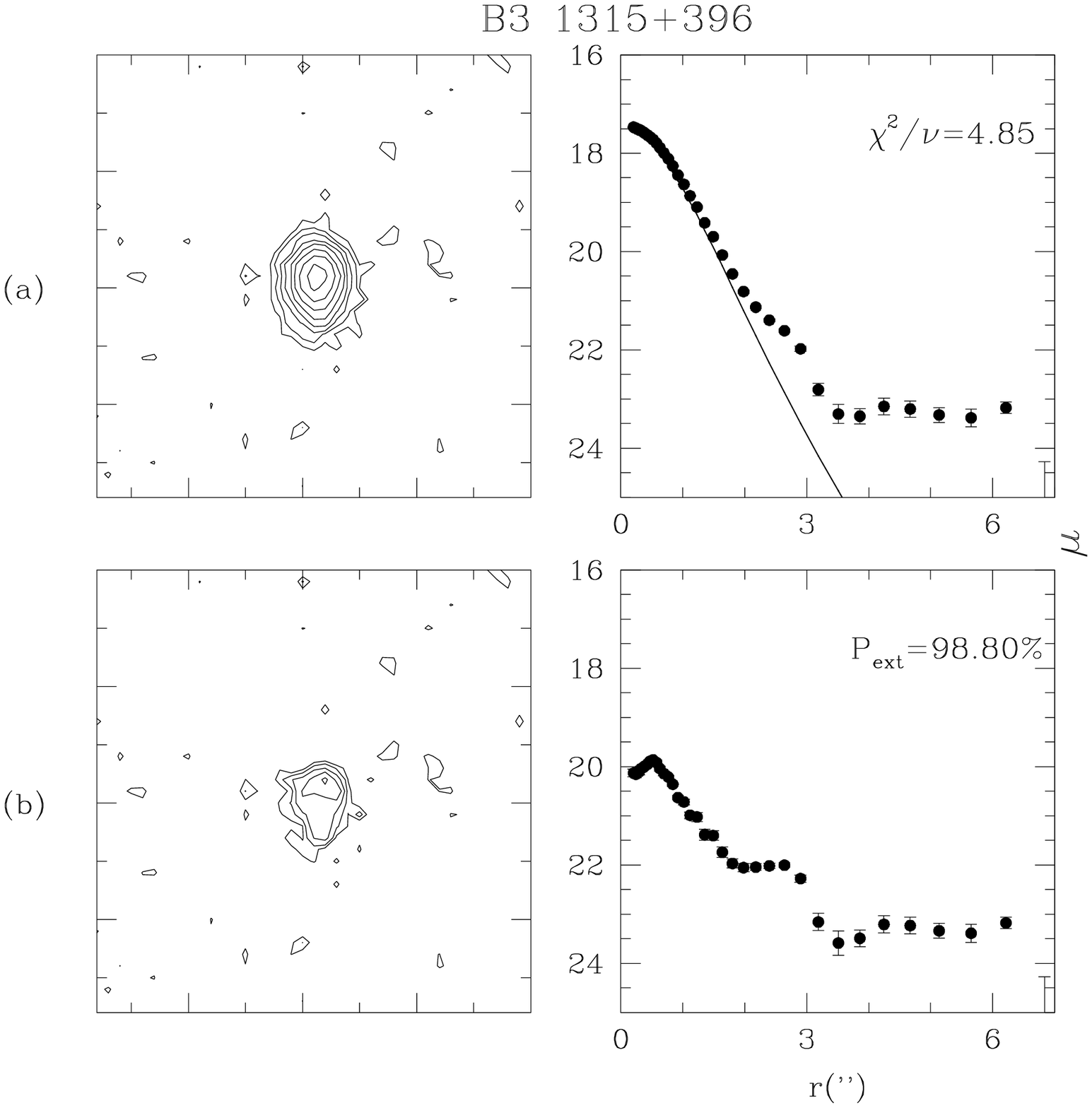}}
\end{minipage}
\epsfxsize=8cm
\begin{minipage}{\epsfxsize}{\epsffile{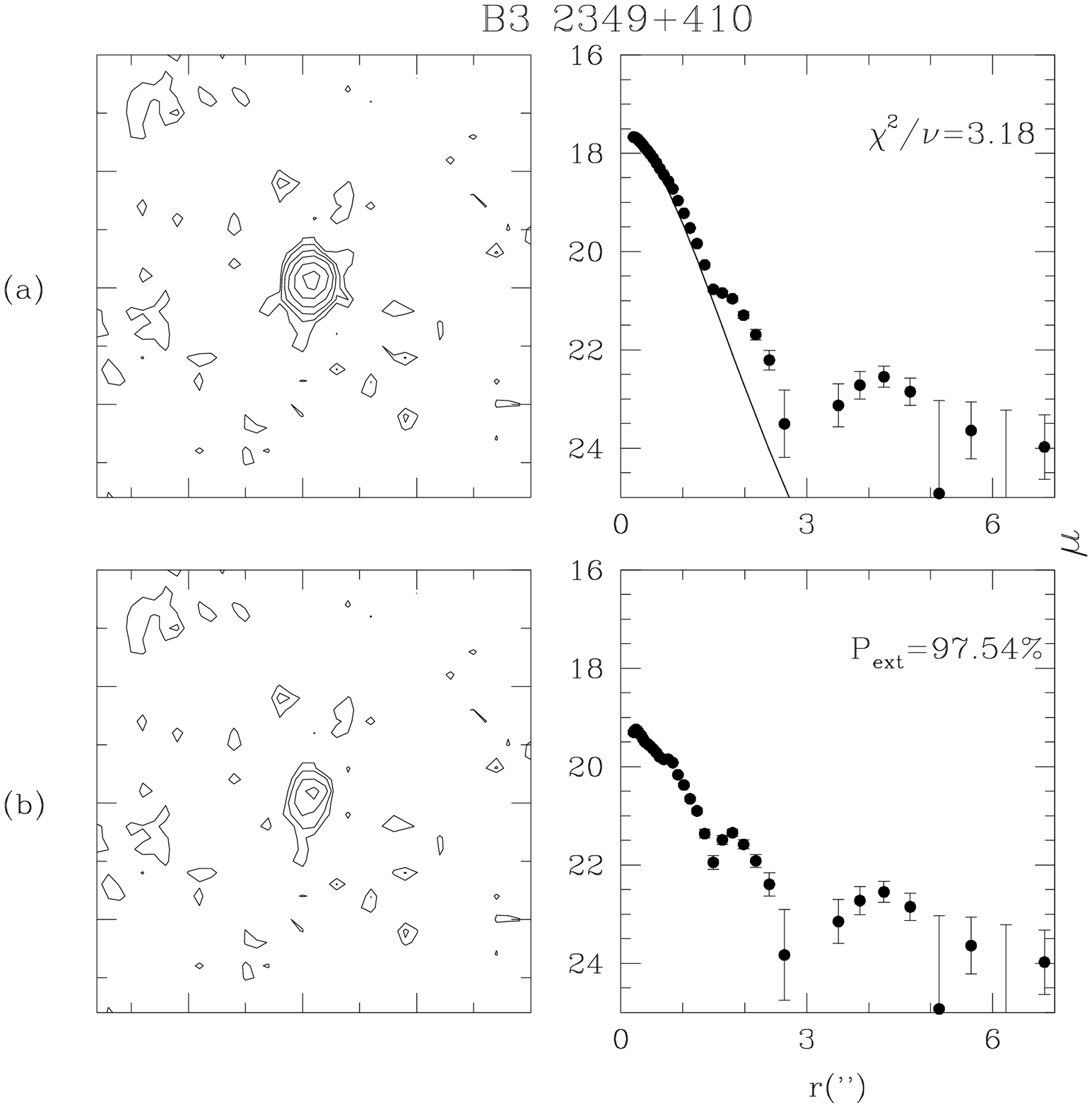}}
\end{minipage}
\epsfxsize=8cm
\begin{minipage}{\epsfxsize}{\epsffile{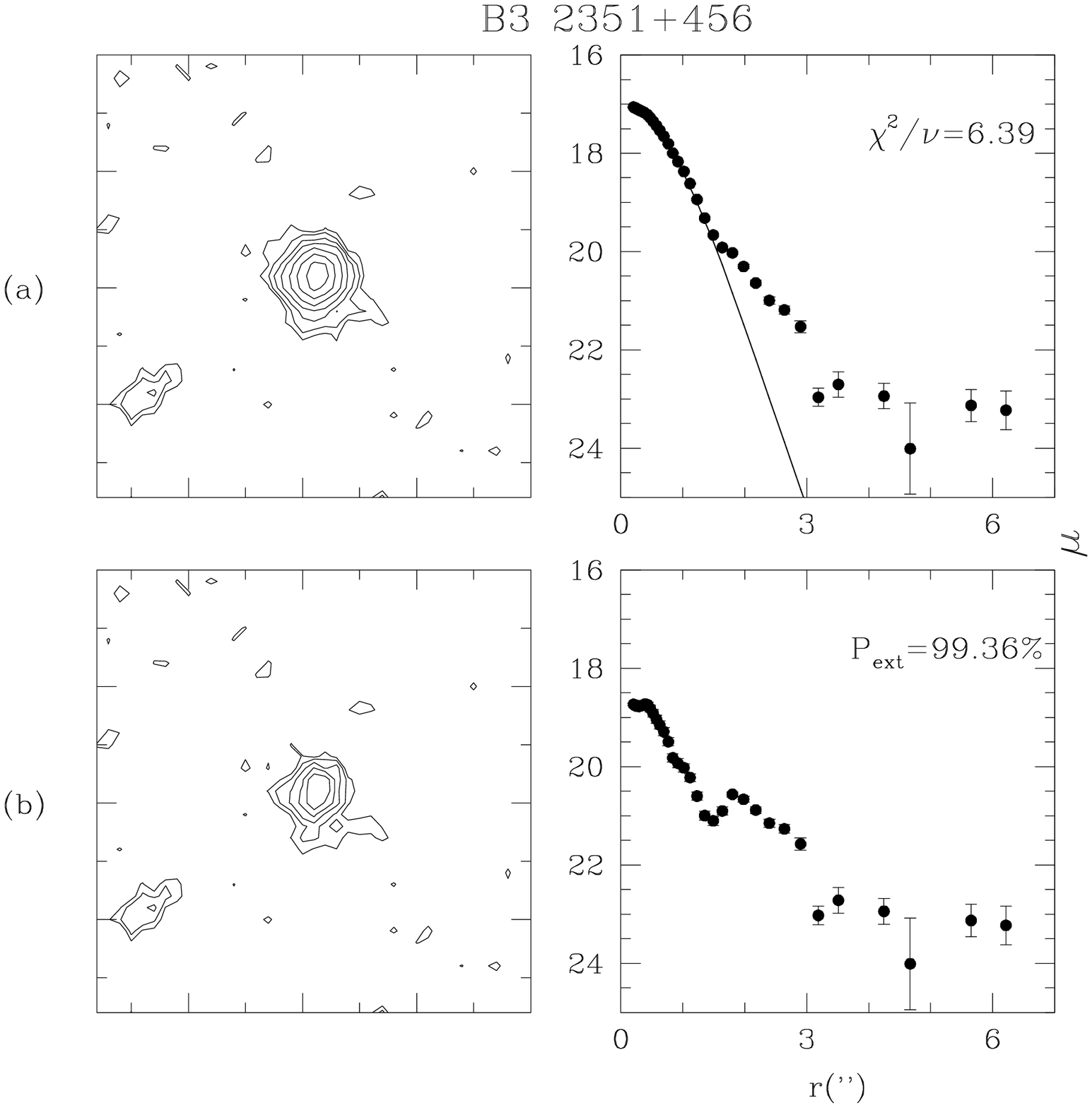}}
\end{minipage}
\end{center}
\caption[]{{\small Continued}}
\label{fig:ima_4}
\end{minipage}
\end{figure*}

%\addtocounter{figure}{-1}

Seven of the 16 objects present distortions, tails or bridges with
possible close companions at $<$30$\arcsec$ (40\% of the sample).
These distortions could be traces of past collisions, interactions or
even merging processes, which affect the host and nearby galaxies.  In
some of them, B3 0219+443 and B3 0740+380C, the optical-to-NIR colours
of the nearby objects ($<$30$\arcsec$) are compatible with the colours
of galaxies at the redshift of the quasars (S\'anchez 2001; S\'anchez
\& Gonz\'alez-Serrano 2002). However, a contamination from a nearby
object, either a galaxy or a faint star, within $\sim$2\arcsec\ of the
quasar could produce a similar effect and create an artificial
distorted HG.  We have estimated the mean density of objects in our
images using the number of objects detected in the February 1999
images (the deepest images with the larger field of view). We found a
density of 2.7$\pm$0.6 10$^{-3}$ obj/arcsec$^2$, including both stars
and galaxies. The large dispersion is due to the fact that these
objects tend to inhabit clusters or groups of galaxies of different
populations (see S\'anchez \& Gonz\'alez-Serrano 2002, also for a
detailed explanation of the procedure to determine the density).
Using these numbers, we expect to find 0.043 objects within a box of
4\arcsec$\times$4\arcsec\ around each quasars. I.e., we expect that a
contaminating source could create an artificial distorted HG in 4.3\%
of the objects. For our sample of 16 extended sources we do not expect
any significant contribution from this effect ($\sim$0.64 possible
contaminted objects).

Together with data from Paper I, we have a collection of 29 HGs
obtained from the analysis of the $K$-band images of 60 radio quasars
from the B3-VLA sample.  Fifteen of them (Paper I) show evidence of
possible interactions ($\sim$50\% of the sample).  This evidence is
tenuous and should be confirmed by spectroscopics studies of both
the host galaxies and the nearby companions.

\section{The properties of the HGs of radio sources}

\subsection{The $K-z$ distribution: The evolution of the HGs}

\begin{figure*}

\centerline{\vbox to 0cm{\vfil}
\epsfxsize=16cm
\epsffile{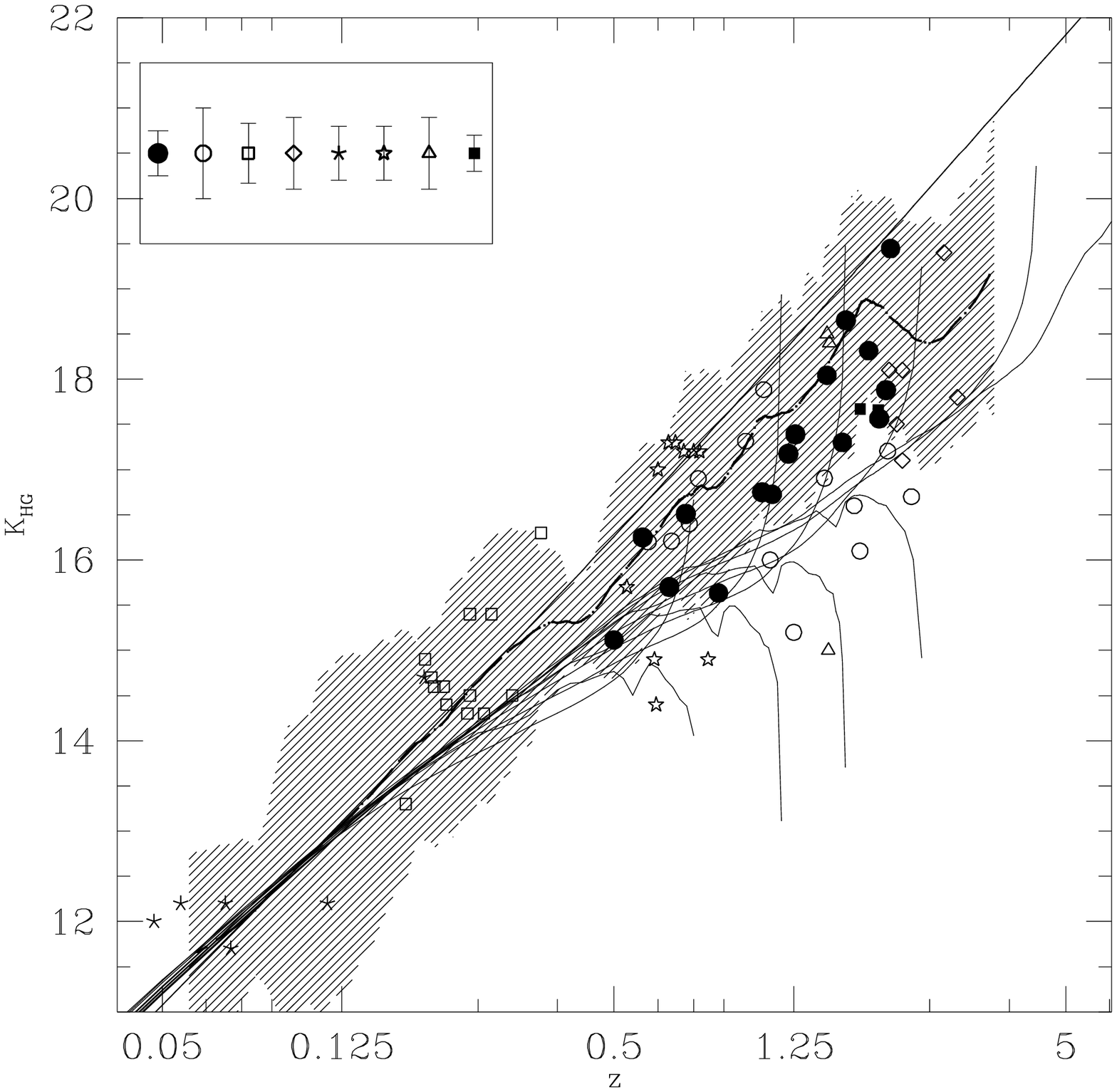}}
\caption[]{{\small 
$K$-band magnitude distribution with redshift
for the HGs of radio quasars: circles for the B3-VLA (solid for new
data and open for Paper I data), diamonds for Lehnert  et al. (1992), open
squares for Taylor et al. (1996), asterisks for Kotilainen et
al. (1998), stars for Kotilainen \& Falomo (2000), triangles for
Falomo et al. (2001) and solid squares for Kukula et al. (2001).
The typical error of the magnitudes for the
different samples is presented in the upper-left panel. The shaded
 region shows the parameter space occupied  by radio galaxies
of different samples (see text); the dashed line is the distribution
of the mean values. The solid concave lines show the expected evolution 
of an elliptical galaxy with different formation redshifts 
($z_{\rm for}$=7,4.5,2.5,1.5,1.0 and 0.7, from right to left).
The convex lines show the expected evolution of an galaxy with an ongoing burst
with different formation redshifts ($z_{\rm for}$=2.5,1.5,1.0 and 0.7, from right to left).
The near straight solid line shows the expected magnitudes for a galaxy without evolution.
}}
\label{fig:k_z}
\end{figure*}

Figure \ref{fig:k_z} shows the $K$-band apparent magnitude of HGs of
radio quasars as a function of redshift. The solid circles represent
the HGs presented in this study, and the open circles the HGs
presented in Paper I. The remaining points represent data obtained
from the literature: diamonds from Lehnert et al. (1992), open squares
from Taylor et al. (1996), asterisks from Kotilainen et al. (1998),
stars from Kotilainen \& Falomo (2000), triangles from Falomo et
al. (2001) and solid squares from Kukula et al. (2001). The HG
magnitudes from Lehnert et al. (1992) were corrected by 0.4 magnitudes
to take into account the underestimation of the HG flux due to the
total subtraction of the central point-like source, following the
indications of the authors. Magnitudes from Kotilainen et al. (1998),
Kotilainen \& Falomo (2000), Falomo et al. (2001) and Kukula et al.
(2001) were transformed from $H$ to $K$ band using the correction
suggested by authors ($H-K\sim$0.2 mag, Kotilainen et al. 1998). We
have used only the 2 HGs detected in the $H$-band from Kukula et al.
(2001) with a reliable result in the radial profile fitting. We have
not considered their detections in the $J$-band (due to the
uncertainties in the transformation to the $K$-band), and their
dubious detections. This comprises a heterogeneous sample of 69 HGs,
29 B3-VLA HGs plus 40 HGs extracted from the literature.

The shaded region shows the space occupied by the radio galaxies of
 different samples ($\pm$1.5$\sigma$ from the mean value), including
 the 3C and B2 (Class 1Jy) samples, studied by Lilly et al.  (1985)
 and Lilly (1989), the 6C and 7C sample, and the HzRG (High-$z$ radio
 galaxies). The distribution of the mean value is represented by the
 dashed line.  All the radio galaxy data have been extracted from
 Fig. 1 of Van Breugel et al. (1999), except for the 7C data, obtained
 from Lacy et al. (2000) and Willot (2002). These radio galaxy
 samples, except for HzRG, are roughly complete (i.e. 3C is 98\%
 complete). In contrast, only low-$z$ quasar host samples are complete
 (e.g. Taylor et al.  1996). The quasar hosts are mainly $\sim$50\%
 complete, as only the brightest objects are detected.

The distributions are roughly similar. Therefore, the HGs of different
families of radio sources have similar luminosities at different
cosmological epochs, down to $z<$3. However, 9 of the 70 HGs presented
in this figure are clearly more luminous than the radio galaxies
($\sim$13\% of the HGs sample), 6 of them at $z>$1.  Considering also
that only half of the HGs of the radio quasars have been recovered (or
even less at $z>1$, as we quoted above), we conclude that although the
mean distributions seem to agree, the dispersion along the mean value
is larger for the HGs. The errors of the HG magnitudes (upper-left
box) are clearly larger than the errors of the magnitudes of the radio
galaxies. This could account for the larger dispersion in the
distribution in Fig. \ref{fig:k_z}. But not all the difference could
be due to this cause: the largest errors were about 0.5 mag, and the
difference between the magnitudes of the high-luminosity HGs and a
brightest radio galaxies at the same redshift was $\sim$1-1.5 mag. We
need to explore other possibilities to explain these extreme
luminosities.

Different evolutionary models are shown in Fig. \ref{fig:k_z}.  Each
solid concave line shows the evolution of an elliptical galaxy with
different formation redshifts ($z_{\rm for}$=7,4.5,2.5,1.5,1.0 and
0.7, from right to left). Each convex line shows the evolution of a
galaxy in a continuous burst process with different formation
redshifts ($z_{\rm for}$=2.5,1.5,1.0 and 0.7, from right to
left). These models have been included to show the most extreme
situations, from pasive to violent evolution, and from old to
young galaxies. The solid near-straight line shows the magnitudes for
a galaxy without evolution, which could be considered the most extreme
situation of a very old galaxy without present star formation. The
evolution tracks have been determined using the GISSEL code (Bruzual
\& Charlot 1993), assuming an intrinsic luminosity of $M_*$ (Mobasher
et al. 1993), and a Salpeter mass distribution with a range of masses
between 0.1 and 125 $M_{\odot}$ (Salpeter 1955). The instrinsic
luminosity has been selected to match the data at low-$z$ for both
families of HGs. We have used the E-type and Burst-type isochrone
spectral energy distributions (ISED) from Bolzonella et al. (2000).

The mean distribution of magnitudes with redshift is well described by
an old E-type galaxy model, near to a no-evolution model.  However,
there is a large dispersion around this mean track.  The dispersion
could be explained by a range of formation redshifts, a range of
different evolutions and/or a range of intrinsic luminosities (see
Figure). In the first two cases, as we explained in Paper I, the
brightest HGs would be galaxies in a violent star formation
process. These galaxies would be blue and their predicted optical
magnitudes would as bright as (or even brighter than) the observed
quasar magnitudes.  Therefore, a large range of star-formation
redshifts or violent  evolution have to be ruled out as a general
explanation to the data.  In the same way, Nolan et al.  (2001) and de
Vries et al. (2000) have presented evidence that low-$z$ quasar hosts
and radio galaxies are dominated by a stellar population of at least
$\sim$12Gyr. It appears more likely that the dispersion in the
distribution is due to a dispersion in the instrinsic luminosity of
the HGs. This dispersion seems to be larger for the HGs of radio
quasars than for radio galaxies.

Eales \& Rawlings (1996), Rawlings (1998), and more recently Willot et
al. (2002) found that the low radio-luminosity radio galaxies (6C and
7C) are fainter at the NIR than powerful ones (3C and B2) at the same
redshift. This implies a correlation between the radio power and the
NIR luminosity of the galaxy. Its effect on the $K-z$ distribution is
an increase in the dispersion. Therefore, the $K-z$ distribution for
the different families of radio sources is well described by a mean
evolution typical for an old E-type galaxy (or a no-evolution model),
and a range of absolute luminosities correlated with the radio
power. The reason for this correlation could be a relation of both
parameters (radio power and HG luminosity) with the central black hole
mass (McLure et al. 1999; Dunlop et al. 2001).

This scenario describes the observed distribution accurately.
Moreover, it explains the narrower distribution observed for the 3C
radio galaxies due to the reduced range of radio luminosities (Lilly
et al.  1985). In this context, we do not find radio galaxies as
luminous in the NIR as the ultraluminous HGs observed here ($\sim$13\%
of the sample) due to a selection effect. However, this scenario
implies a relation between the HG luminosities and the radio power.

\subsection{Relation between the HG luminosity and the radio power}

\begin{figure}
\centerline{\vbox to 0cm{\vfil}
\includegraphics[angle=270,width=9cm]{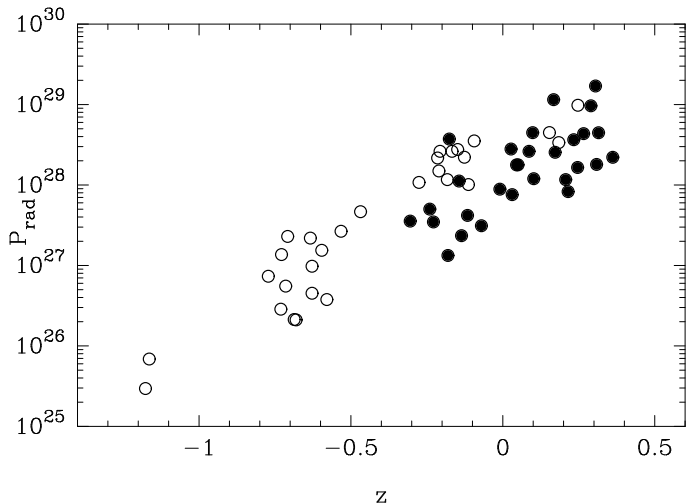}
}
\caption[ Distribution of the radio power at 1460~MHz against the
redshift. The solid symbols are the HGs for the B3-VLA quasars; 
the open symbols are the HGs from the literature.
]
{ Distribution of the radio power at 1460~MHz against the
redshift.  The solid symbols are the HGs for the B3-VLA quasars; 
the open symbols are the HGs from the literature.}
\label{fig:da_z3}
\end{figure}

Figure \ref{fig:da_z3} shows the radio power of the 29 HGs of B3-VLA
quasars plus the 30 HGs obtained from literature with published flux
at 1460~MHz (out of the original 40 HGs). The radio power has been
obtained asuming a spectral index of $\alpha$=$-$1 for the radio
emission, in all the objects. There is a clear correlation between
these parameters for both the heterogeneous sample
($r$=0.81;$P>$99.99\%), and the B3-VLA sample ($r$=0.70;$P$$=$99.88\%).
The B3-VLA subsample covers a range of radio-power/redshifts not very
well covered before by data in the literature, extending the studies to
more powerful/higher redshifts objects.

\begin{figure}
\centerline{\vbox to 0cm{\vfil}
\includegraphics[angle=270,width=9cm]{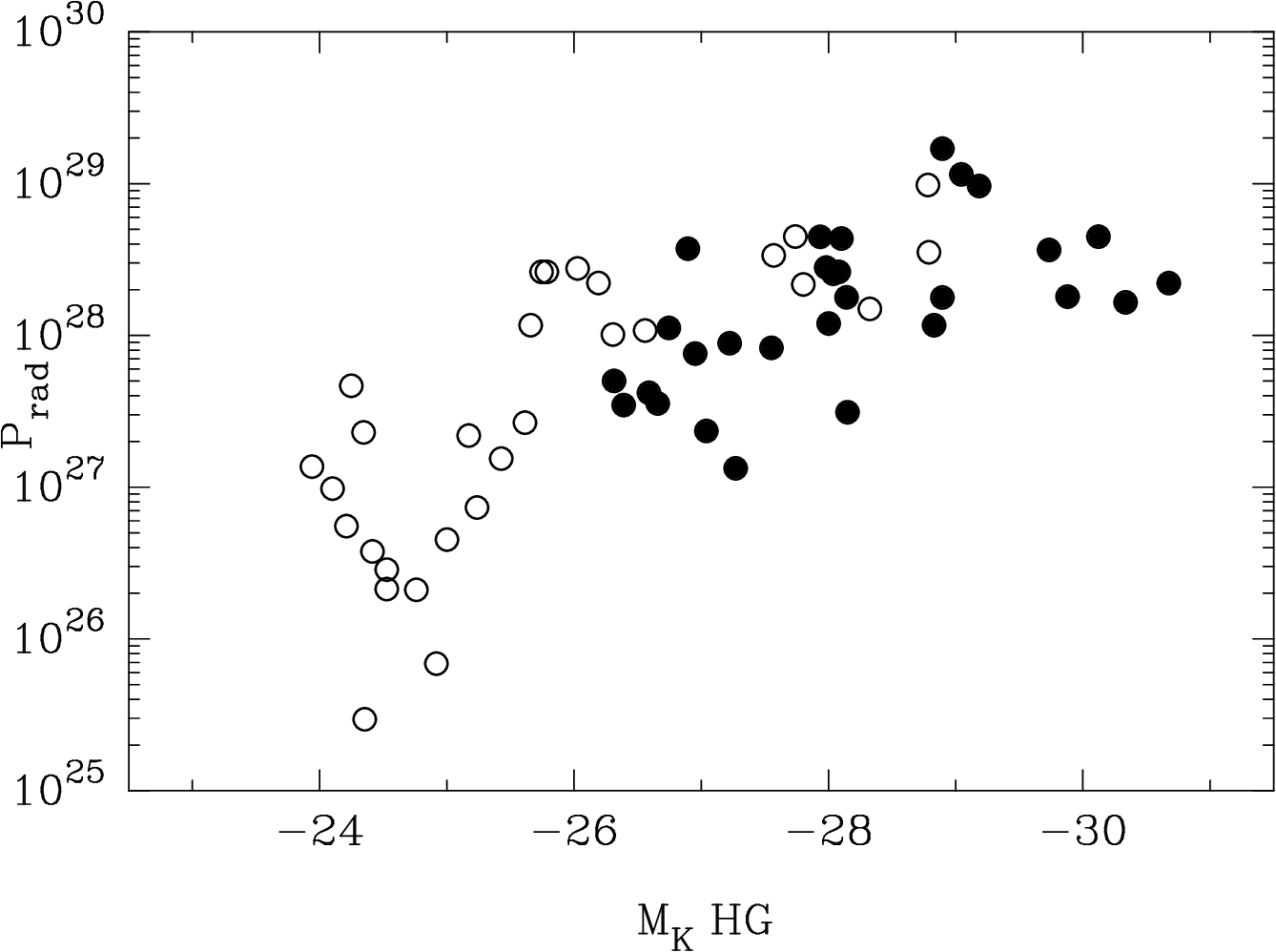}
}
\caption[
]{{\small Distribution of the radio power at 1460~MHz against the
absolute magnitude of the HGs.  The solid symbols are the HGs for the B3-VLA quasars; 
the open symbols are the HGs from the literature.
}}
\label{fig:da3}
\end{figure}

Figure \ref{fig:da3} shows the distribution of radio power at 1460~MHz
against the absolute magnitude of the HGs. These parameters are
clearly correlated, both for the heterogeneous sample
($r$=-0.74;$P>$99.99\%) and the B3-VLA subsample
($r$=-0.56;$P$$=$99.81\%). This correlation can be accounted for by
the strong dependence on redshift shown by both parameters (the
absolute magnitude of the host galaxies correlates with redshift,
$r$=-0.60;$P$$>$99.99\%). Splitting the sample by the mean redshift,
we find that there is still a correlation for the low-$z$ subsample
($r$=-0.61; $P>$99.99\%; $n_{\rm gal}$=37), although only a faint
tendency for high-$z$ subsample ($r$=-0.32, $P$=83.93\%; $n_{\rm
gal}$=22). It seems that there is an intrinsic relation between these
two parameters, not induced by their depencence with redshift.  Similar
tendencies have been described for radio galaxies (Ledlow \& Owen
1996, Rawlings et al.  1998; Lacy et al.  2000; Jarvis et al. 2001;
Inskip et al. 2002; Willot et al.  2002), and radio-quiet quasars, of
which the HGs at high $z$ ($z$$\sim$2-3) are 2 mag fainter than
powerful radio galaxies at the same redshift (Rawlings et al. 2001).

There is a simple explanation of why we may expect to find that
stellar luminosities of HGs are correlated with radio luminosity. In
essence, this comes down to the fact that more massive objects are
generally more luminous. In this case there are several observed
correlations that can be used to specify the relationship between
stellar and radio luminosity. First, the stellar luminosity of
ellipticals is known to correlate almost linearly with the central
black hole mass (Magorrian et al. 1998), presumably because the
stellar luminosity is very closely related to the stellar mass and there
is a tight correlation between host and black hole masses (Gebhart et
al. 2000; Ferrarese \& Merrit 2000). The radio luminosity is found to
be linearly related to the narrow emission line luminosity (Willot et
al. 2001) and therefore to the UV ionizing luminosity (Rawlings \&
Saunders 1991; Willot et al. 1999; Carballo et al. 1999). The UV
ionizing luminosity is due to the accretion of material
onto the supermassive black hole. In fact, Franceschini et al. (1998)
found a direct correlation between the radio luminosity and the mass
of the central black hole, for a sample of low redshift galaxies.
Therefore any correlation between stellar and radio luminosity is
likely to have its cause in the fact that both parameters correlate
positively with the black hole mass. McLure et al. (1999), and more
recently Dunlop et al. (2001), reached a similar conclusion studying
a sample of low-$z$ HGs. Our result extends theirs on the
radio-loud sources to a wide range of redshifts (0$<z<$3), increasing
the statistical significance.

\subsection{Relation between the luminosities of the HG and the central
source}

There are several pieces of evidence supporting a scaling of the nuclear
luminosity with HG luminosity. As we quoted above, Magorrian et al
(1998) found a correlation between the HG luminosity and the central
black-hole mass. Since the accretion of matter is the main engine of
the nuclear activity, it is expected that nuclear luminosity scales
also with the black-hole mass. Indeed, Dunlop et al.  (2001) have
shown that radio-loud quasars emit within a range of a 1-10\% of the
Eddington luminosity. This luminosity is proportional to the
black-hole mass. Therefore, a relation is expected between HG and
nuclear component luminosities. On the other hand, hierarchical galaxy
formation models of Kauffman \& Haehnelt (2000), in which they have
addressed both the formation of bulges/elliptical galaxies and the
formation and fueling of their associated black holes, predict a faint
correlation between both luminosities.

\begin{figure}
\centerline{\vbox to 0cm{\vfil}
\includegraphics[angle=270,width=9cm]{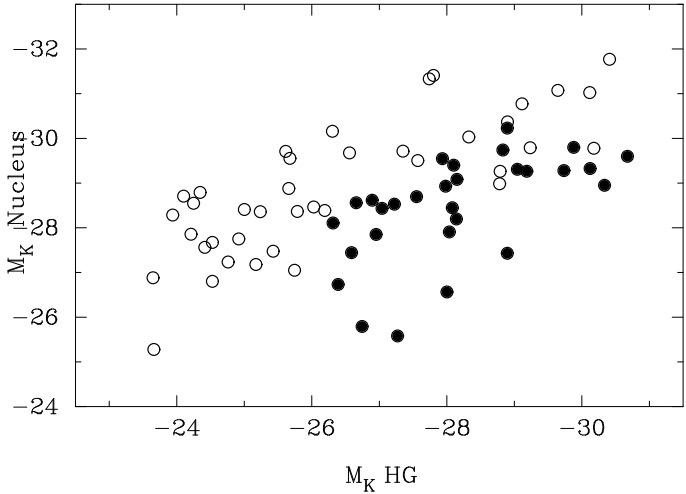}
}
\caption[]{{\small Distribution of the $K$-band absolute magnitudes of the HGs
against the absolute magnitude of the central source.  The solid symbols are the HGs for the B3-VLA quasars; the open symbols are the HGs from the literature.}}
\label{fig:qso_gal}
\end{figure}

\begin{figure}
\centerline{\vbox to 0cm{\vfil}
\includegraphics[angle=270,width=9cm]{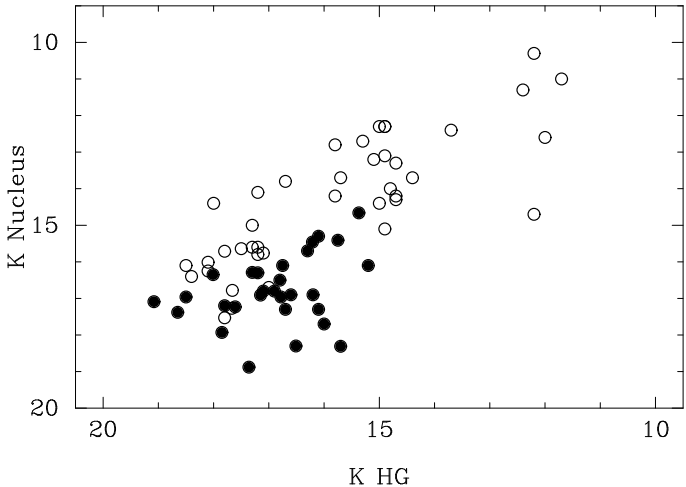}
}
\caption[]{{\small Distribution of the $K$-band apparent magnitudes of the HGs
against the apparent magnitude of the central source.  The solid symbols are the HGs for the B3-VLA quasars; the open symbols are the HGs from the literature.}}
\label{fig:qso_gal_obs}
\end{figure}

Figure \ref{fig:qso_gal} shows the distribution of the absolute
magnitudes of the HGs against the absolute magnitude of the nuclear
source. There is a clear correlation for the heterogeneous sample
($r$=0.56;$P$$>$99.99\%) and the B3-VLA subsample
($r$=0.57;$P$$=$99.88\%). The strong depence that both parameters show
with redshift could contribute significantly to this correlation (the
absolute magnitude of the central sources also correlates with
redshift, $r$=-0.89;$P>$99.99\%). We have determined the correlation
coefficient for the low and high-$z$ subsamples, finding that there is
still a correlation in both subsamples ($r$=0.45;$P$=99.61\% and
$r$=0.41;$P$=97.37\%, respectively). In order to remove the redshift
effect we have determined the correlation coefficients for the
apparent magnitudes, instead of the absolute magnitudes. Figure
\ref{fig:qso_gal_obs} shows the distribution of the apparent
magnitudes of the HGs against the apparent magnitude of the nuclear
source.  We have found that the correlation is even stronger for the
heterogeneous sample ($r$=0.74;$P$$>$99.99\%), although only a slight
correlation is found for the B3-VLA subsample, which covers a narrower
range of apparent magnitudes ($r$=0.34;$P$$=$96.79\%).  An intrinsic
relation between both parameters rather than a redshift induced
relation seems to be the correct explanation for the observed
distributions.

The distributions in Figs \ref{fig:qso_gal} and \ref{fig:qso_gal_obs}
could be induced by different biases.  The QSO selection criteria as
point-like sources could depopulate the lower-right region (bright HGs
harboring a faint nuclear source). For the B3-VLA sample, the
point-like criterion was applied to POSS plate data (Vigotti et al.
1997). Similar data were used for the different samples included in
our heterogeneous sample (e.g.  Drinkwater et al. 1997 for the PKS).
Taking into account the mean $B-K$ colours of the quasars and the
elliptical galaxies ($\sim$2.5 and $\sim$5 mag respectively), we do
not expect a significant effect of this bias on the observed
distribution.  E.g., a quasar with $B$$=$17 mag, classified as
point-like using the POSS plates could have a host galaxy that
contributes 50\% to its $K$-band flux, and it would be undetectable on
POSS plates ($B_{\rm HG}\sim$20.5 mag).

On the other hand, the upper-left region (faint HGs harboring a bright
nuclear source) could be depopulated due to the incompleteness in the
HG detection. Only HGs that contribute more than 10\% to the total
integrated flux are considered as real detections. This imposes an
upper limit envelope on the observed distributions near to $K_{\rm
gal}$=$K_{\rm nuc}$+2.5. If this is the case we would expect an
increase of the strength of the correlation with increasing
incompleteness.  However, as stated before, this strength is the same
for the low-$z$ and high-$z$ subsamples, but the completeness
decreases significantly ($\sim$75\% and $\sim$40\% respectively, for
the B3-VLA sample). Moreover, this argument could only be applied to
the B3-VLA data. The heterogeneous sample was built using HG samples
that are mainly complete or less affected by uncompleteness (due
mainly to the lower redshift range they cover). Indeed, the
distribution of B3-VLA data in Fig. \ref{fig:qso_gal_obs} does not
show a clear envelope at the predicted 2.5 mag from the K$_{\rm
Nucleus}$=K$_{\rm HG}$ line.

These correlations confirm the tendency presented in previous articles
for more powerful quasars to reside in more luminous hosts (e.g.,
Paper I, Kotilainen et al. 1998, Kotilainen \& Falomo 2000, Taylor et
al. 1996). As we stated above, the most probable reason for this
correlation is a relation of both luminosities with central black
hole mass. This correlation is weaker or even absent in the more
nearby, lower luminosity and radio-quiet AGNs (e.g. McLeod \& Rieke
1995; McLeod \& Rieke 1994a,b; Percival et al.  2001). The smaller dynamical
range of magnitudes covered by these samples could account for the absence of 
this correlation.

\subsection{Structural parameters of the HGs}

The mean absolute magnitude of the HGs is $\sim$-27.02$\pm$1.81 mag,
which corresponds to a luminosity of $\sim$3L$_{*}$ ($M_{K}^*$=-25.86
from Mobasher et al. 1993). $\sim$75\% of the objects are brighter
than $L_{*}$. Their mean effective radius is 15.00$\pm$13.90 kpc and
all but one have $r_{\rm e}$$>$3.5 kpc.  Capaccioli et al. (1992) found that
there is a limit to the size of the galaxies ($\sim$4.5 kpc), below
which it is rare to find AGNs, BCG and/or cDs. As expected, $\sim$90\% 
of the HGs of our sample are above this limit. They are
large and luminous galaxies, similar to BCGs or cDs, which are in the
range of the brightest known galaxies.

Elliptical galaxies show fundamental relations between their
luminosities, their effective radii and their velocity dispersions
(Hamabe \& Kormendy 1987; Mobasher et al. 1999).  These relations
define a plane in the three-dimensional space described above, the
plane known as the Fundamental Plane (FP). The projection of this
plane on the luminosity-radius plane ($\mu_{\rm e}-r_{\rm e}$)
determines a relation between both parameters (Hoessel et al. 1987),
with a slope near to $\sim$3. This relation, of which the origin is
still unknown, is shown by all the elliptical galaxies, and limits the
structural models presented for these objects. It has been suggested
that episodes of collision/merging could produce this relation
(Capaccioli et al.  1992).

\begin{figure}
%\vspace{8.0cm}
\centerline{\vbox to 0cm{\vfil}
\epsfxsize=9cm
%\epsffile{k_z2.ps}}
\epsffile{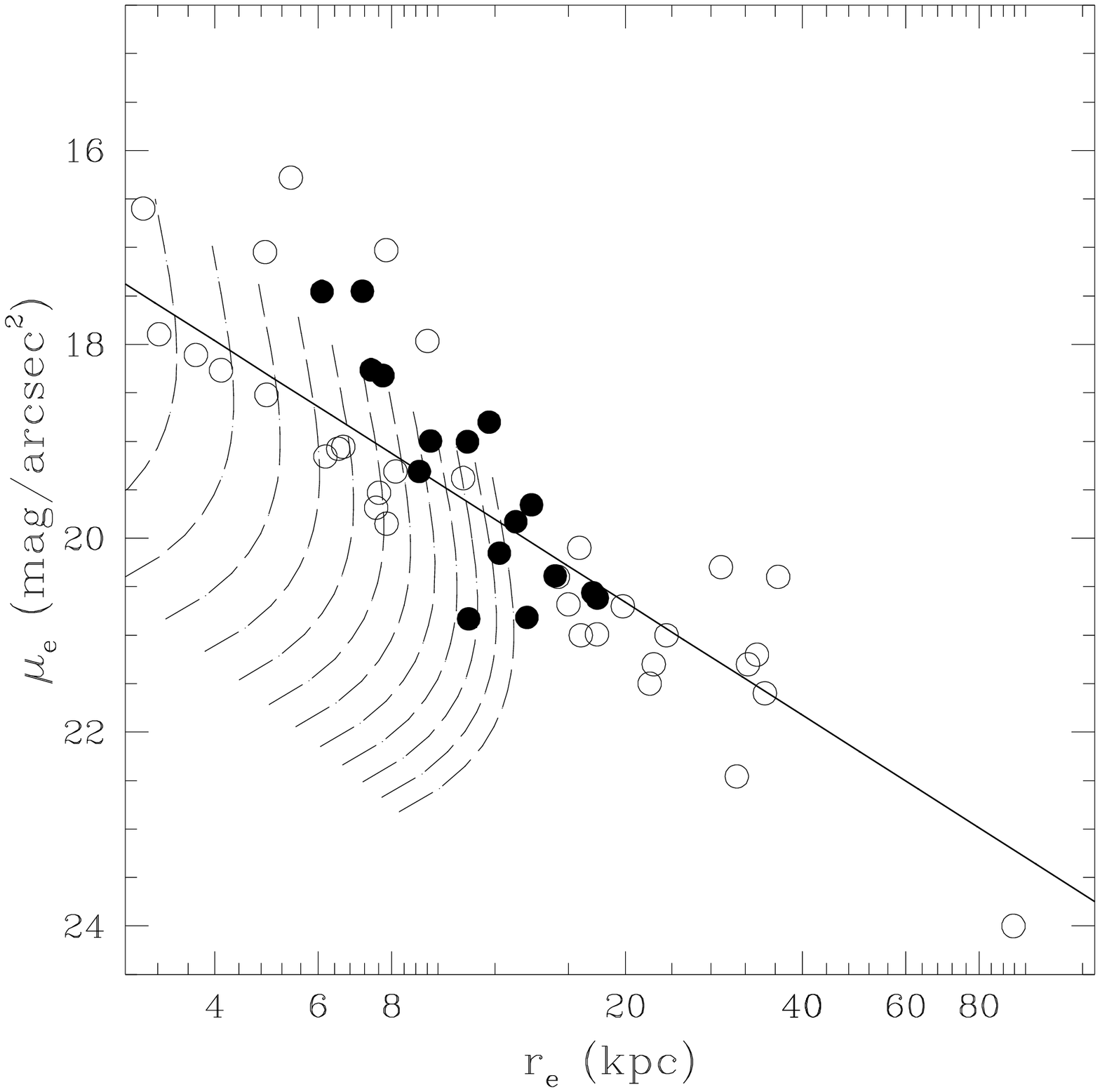}}
\caption[
]{{\small Distribution of the effective surface brightness profile
    against the effective radius. The solid symbols are the HGs for
    the B3-VLA quasars; the open symbols are the HGs from
    the literature. The solid line show the best linear regression fit to the
    unbiased data. The dashed lines show the exclusion region for detection for the  B3-VLA subsample.
}}
\label{fig:k_z2}
\end{figure}

We have determined the effective surface brightness of the HGs
assuming a $r^{1/4}$ profile, correcting for {\it cosmological
  dimming} (e.g., Kolb \& Turner 1990).  Figure \ref{fig:k_z2} shows
the distribution of the effective surface brightness with
effective radius for the 50 HGs of our sample with published effective
radius.  There is a clear correlation between both parameters, with a
slope slightly larger than 3 ($\sim$4.6), roughly similar to the value
reported by Taylor et al. (1996). Our sample is biased towards bright
and large galaxies, especially at high redshift.  The lack of small
and faint galaxies could artificially increase the slope of the
correlation. The dashed lines show the location on this plane of a
galaxy with constant magnitude ($K$$=$19 mag) and constant effective
radius at different redshifts. Each line corresponds to a certain
effective radius within the range
0.1$\arcsec$$<$$r_{\rm e}$$<$1.5$\arcsec$. They represent the region of
parameters where the uncompleteness of the detection rise up to
$>80$\% for the B3-VLA data (solid circles).  Similar exclusion
regions could be defined for each sample plotted in the figure.  We
have restricted the linear regression analysis to the data less
affected by this bias ($r_{\rm e}>$10 kpc, $n$=27 HGs). We found
a strong correlation ($r$=0.80, $P>$99.99\%), of the form:
$$\mu_{\rm e} = 15.64_{\pm 0.60}+3.86_{\pm 0.46} log(r_{\rm e})$$ similar to the
well-known relation for the elliptical galaxies.

\section{Relation between the absolute magnitude of the nuclear source
and the radio power}

\begin{figure}
\centerline{\vbox to 0cm{\vfil}
\includegraphics[angle=270,width=9cm]{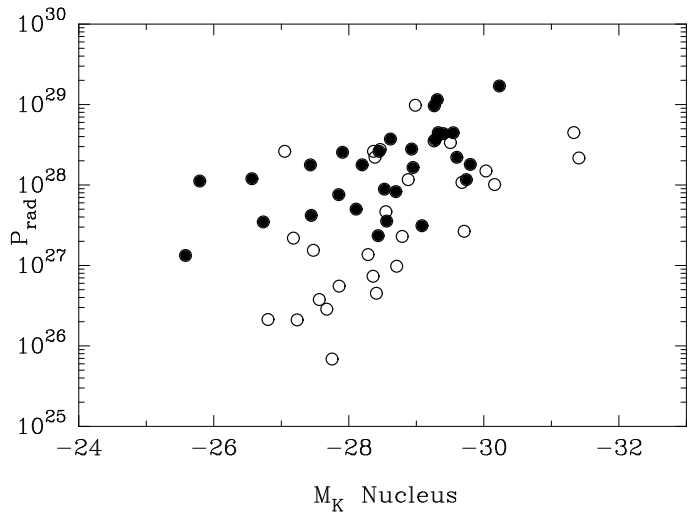}
}
\caption[
]{{\small Distribution of the radio power at 1460~MHz against the
absolute magnitude of the nuclear source.  The solid symbols are the HGs for the B3-VLA quasars; the open symbols are the HGs from the literature.
}}
\label{fig:da1_2_lpr}
\end{figure}

Figure \ref{fig:da1_2_lpr} shows the distribution of the radio power
of the quasars against the $K$-band absolute magnitude of the nuclear
source. Both parameters are clearly correlated for the heterogeneous
sample ($r$=0.57;$P>$99.99\%) and the B3-VLA sample
($r$=0.58;$P=$99.88\%). This correlation could be induced by the
dependence of both parameters on redshift. Splitting the sample in
low-$z$ and high-$z$ subsamples by the mean redshift, we found that
the correlation persists for both subsamples ($r$=0.52,$P$=99.99\% and
$r$=0.51,$P$=98.20\% respectively). This suggests that it is not
induced by the redshift. The derived relation between the $K$-band
luminosity of the nuclear source and radio power is $L_{\rm
K,qso}\propto P_{\rm 1460MHz}^{0.37\pm0.04}$. Carballo et al. (1999)
found a similar relation between the optical-UV luminosity and radio
power at 408 MHz, $L_{2400\AA}\propto P_{\rm 480MHz}^{0.52\pm0.10}$,
for the B3-VLA quasars.  These correlations imply a nearly one-to-one
relation between the optical and NIR emission, which suggests that the
principal mechanism of both kind of emissions is the same.  The
correlation between optical-NIR and radio emission in quasars has been
discussed in Rawlings \& Saunders (1991), Serjeant et al.  (1998),
Willot et al.  (1999), Carballo et al. (1999) and Simpson \& Rawlings
(2000). It is outside the scope of the present study to discuss it in
detail.

\section{Discussion and Conclusions}

We have presented in this study the $K$-band images of 31 radio
quasars from the B3-VLA sample, with a redshift range between
$z\sim$0.5 and $z\sim$3.2. A new procedure, based on the analysis of
the surface brightness profile, has been proposed to estimate the flux
contributions of both the  point-like nucleus and the HG of the
quasars. This procedure, tested on a wide sample of simulated images
and field stars, makes it possible to assign a probability of being extended to
each quasar, recover the flux and effective radius of the HG, classify
it as elliptical or spiral, and restore the image of the HG. The
simulations and tests on field stars set limits to the detection and
reliability of the recovered parameters. A different procedure, used
in Paper I (Carballo et al. 1998), has also been tested on the
simulations, in order to allow the use of the combined data.

We have detected the HGs of 16 of the 31 quasars ($\sim$50\% of the
sample), all of them with a probability of being extended of higher than
95\%, and 10 of more than 99\%. Together with data from Paper I, we have a
sample of 29 HGs of B3-VLA quasars. Searching the literature, we
have built a sample of heterogeneous HGs of radio-loud quasars that
comprises 69 HGs. These HGs are large ($r_{\rm e}\sim$15 kpc) and
luminous ($\sim$75\% of them brighter than $L_*$) elliptical
galaxies. They follow the $\mu_{\rm e}$-$r_{\rm e}$ relation for elliptical
galaxies. About $\sim$45\% of the B3-VLA HGs show distortions and
tails in their images that could be evidence of a merging/colliding
process.  This result agrees with the hypothesis that considers a
merging process as the origin or triggering mechanism of nuclear
activity (e.g. Smith \& Heckman 1990; Hutching \& Neff 1992).  There
is other evidence that supports this scenario: (i) the quasars are
located in overdensities of galaxies compatible with being at the same
redshifts (e.g. S\'anchez \& Gonz\'alez-Serrano 1999; S\'anchez 2001),
and in a spatial position where a collision is more likely to
produce a merger (S\'anchez \& Gonz\'alez-Serrano 2002); (ii) the
traces of recent star formation processes induced by interactions (e.g.
Nolan et al 2001, Canalizo \& Stockton 2000); (iii) the need of a
violent process that produces an infall of mass in the inner region to
explain nuclear activity (e.g.  Bekki \& Noguchi 1994).  

However, it is not clear that the fraction found could be considered
as a significant excess of merger galaxies, even if all of them were
confirmed. The fraction of merger/irregular galaxies in the local
universe is $\sim$7\%, based on classical studies like the
Shapley-Ames Catalog (Shapley \& Ames 1932). But this fraction is
known to increase with redshift. E.g, van den Bergh et al. (1996) and
Lee \& Hwang (2000) found that $\sim$39\% of the galaxies on the
HDF-N and HDF-S show irregular/merger morphologies. We do not have a
proper sample of inactive galaxies with a similar redshift range and
detailed morphological classification to known if the fraction of
mergers found is or IS not a significant excess.

All the HGs are elliptical galaxies morphologically similar to radio
galaxies. The mean luminosity evolution along a wide range of
redshifts (0$<z<$3) is also similar to the evolution shown by radio
galaxies. This evolution is well described by a model with a single
burst of star formation at very high redshift ($z_{\rm for}>$7) and
little or no evolution since that period. It seems that the galaxies
have been formed beyond $z\sim$3, without major burst process down to
this redshift. This supports the unification schemes of radio sources,
since both families of radio sources are similar in features that do
not depend on the angle-of-view.

The HGs present a larger dispersion in mean evolution compared with
the dispersion found in radio galaxies, especially beyond $z>$1.  This
dispersion is most probably due to a correlation between the absolute
magnitude of the HGs and the radio power. Therefore, the most powerful
radio quasars inhabit the most luminous HGs. The absolute magnitudes
of the host and nuclear source (and this last parameter and radio
power) are also correlated. These correlations are most probably due
to a correlation of all three parameters with the central black hole
mass (McLure et al 1999). Dunlop et al. (2001) found that this
tendency is seen both in radio-loud and radio-quiet samples, which
could imply that the same mechanism produces the radio emission in
both kinds of AGN.

In a toy scenario, succesive merging processes could increase the mass
of the spheroidal component of the HG, increasing its luminosity.
These merging processes would feed the nuclear AGN, increasing also
the mass of the central black hole. This will also increase the
nuclear luminosity and radio power. This scenario is consistent with
the result that indicates that radio-quiet quasars inhabit less-populated 
areas than radio-loud quasars (Hall et al.  1998). S\'anchez
\& Gonz\'alez-Serrano (1999) and S\'anchez \& Gonz\'alez-Serrano
(2002) have found that QSOs inhabit the outer regions of clusters
(instead of the core) where a collision is more likely to produce a
merger. Recent morphological analysis of HGs of different families of
AGNs at $z$$<$0.3 found no traces of recent merging processes (Dunlop
et al. 2001). This could indicate an evolution in the merging rate,
although much more detailed studies are needed before a
conclusion can be reached.

The global scenario suggested by the results presented in this article
has to be tested. The first obvious step would be to increase the
fraction of detected HGs in order to reduce the biases, and to extend
the studies to other samples with higher and lower radio
powers. Photometric studies of the HGs at other wavelengths, and
spectroscopy of them and their near companions are needed to determine
better the evolution of these galaxies, and search for possible traces
of merging processes.

\begin{acknowledgements}

We thank the anonymous referee for his valuable and interesting comments
that has increased the quality of this article. We would also thank Jet
Katgert (A\&A Deputy Editor), for his correction of the poor quality english
of the first version of this article.

We thank L.Cay\'on and I.Ferreras for making available the GISSEL
code, and for their support with its operation. We thank all Calar
Alto observatory staff for the friendly support and the inestimable
help in the successive observing runs.  S.F.S\'anchez thanks Danny
Lennon (ING Head of the astronomy group) for their support in the
realization of this research.  S.F.S\'anchez thanks Mariano Tejedor
(Eresmas SA) for his kind help in letting him time and computers to
work on this project.  This project has been partially funded by
grants from the Spanish Ministerio de Educacion y Cultura and
Ministerio de Ciencia y Tecnologia, with the projects PB98-0409 and
AYA2002-03326, respectively.  This project has be partially founded by
the Euro3D Training Network on Integral Field Spectroscopy, funded by
the European Commission under contract No. HPRN-CT-2002-00305.

3.5m CAHA Visiting Astronomer, German-Spanish Astronomical Centre,
Calar Alto, operated by the Max-Planck-Institute for Astronomy,
Heidelberg, jointly with the Spanish National Commission for
Astronomy.

Based on observations made with the WHT operated on the island of La
Palma by the Isaac Newton Group in the Spanish Observatorio del Roque
de los Muchachos of the Instituto de Astrofisica de Canarias.  The
authors want to thank M.Bolzonella for his help with the GISSEL code
and the isochrone spectral energy distribution data.  The authors want
to thank C.R.Benn for HIS inestimable help. His comments, tests and
objective criticism concerning the data have strongly influenced the
final version of the article.

\end{acknowledgements}


\begin{thebibliography}{}

\bibitem[Antonucci 1993]{ant93} Antonucci R., 1993, ARAA, 31, 473

\bibitem[Aretxaga et al. 1998]{are98} Aretxaga I., LeMignant D.,
Melnick J., Terlevich R.J., \& Boyle B.J., 1998, MNRAS, 298, 13

\bibitem[Bahcall et al. 1994]{bah94} Bahcall J.N., Kirhakos S., \&
Schneider D.P.,1994, ApJ, 435,L11

\bibitem[Bahcall et al. 1995a]{bah95a} Bahcall J.N., Kirhakos S., \&
Schneider D.P.,1995a,ApJ,447, L1 

\bibitem[Bahcall et al. 1995b]{bah95b} Bahcall J.N., Kirhakos S., \&
Schneider D.P.,1995b,ApJ,450,486

\bibitem[Bahcall et al. 1996]{bah96} Bahcall J.N., Kirhakos S., \&
Schneider D.P.,1996,ApJ,457,557

\bibitem[Bahcall et al. 1997]{bah97} Bahcall J.N., Kirhakos S.,
Saxe D.H., \& Schneider D.P.,1997,ApJ,479

\bibitem[Bekki \& Noguchi 1994]{bn94} Bekki K., \& Noguchi M., 1994, A\&A, 290, 7

\bibitem[Bolzonella et al. 2000]{bol00} Bolzonella M., Miralles J.M., \& Pell\'o R., 2000, A\&A, 363, 476

\bibitem[Boroson \& Oke 1982]{bo82} Boroson T.A., \& Oke J.B, 1982,
Nature, 296, 3\
97

\bibitem[Boroson \& Oke 1984]{bo84} Boroson T.A., \& Oke J.B., 1984, ApJ,
281, 535

\bibitem[Bruzual \& Charlot \ 1993]{bru93} Bruzual A.G.,\& Charlot S.
, 1993, ApJ, 405, 558

\bibitem[Canalizo \& Stockton 2000]{CS2000} Canalizo G., \& Stockton A.,
2000, ApJ, 328, 201

\bibitem[Canalizo \& Stockton 2000]{CS2000} Canalizo G., \& Stockton A.,
2001, ApJ, 555, 719

\bibitem[Capaccioli et al. 1992]{cap92} Capaccioli M., Caon N., \& D'Onofrio M., 1992, MNRAS, 259, 323

\bibitem[Carballo et al. \ 1998]{car98} Carballo R., S\'anchez S.F.,
Gonz\'alez-Serrano J.I., Benn C.R., \& Vigotti M., 1998, AJ 115, 1234

\bibitem[Carballo et al. \ 1999]{car99} Carballo R., Gonz\'alez-Serrano J.I, Benn C.R, S\'anchez S.F., \& Vigotti M., 1999, MNRAS, 306, 137

\bibitem[Casali \& Hawarden 1992]{cas92} Casali M.M., \& Hawarden T.G., 1992, UKIRT Newsletter, 4, 33

\bibitem[Courbin et al. 2002]{cour02} Courbin F.  Letawe  G., Magain  P., et al., 2002, A\&A, 394, 863

\bibitem[de Vaucouleurs \ 1948]{dva48} de Vaucouleurs G., 1948, Ann. Astrophys., 11, 247

\bibitem[de Vries etal \ 2000]{dvr00} de Vries W.H., O'Dea C.P., Barthel P.D., et al.,
2000, AJ, 120, 2300

\bibitem[Disney et al. \ 1995]{dis95} Disney M., Boyce P.J., Blades J.C.
, et al., 1995, Nature 376, 150

\bibitem[Drinkwater et al. \ 1997]{dri97} Drinkwater, M. J., Webster, R. L., Francis, P. J., et al., 1997, MNRAS, 284, 85  

\bibitem[Dunlop \& Peacock 1993]{dunPea93} Dunlop J.S., \& Peacock J.A., 1993, MNRAS, 263, 936

\bibitem[Dunlop et al. 1993]{dun93} Dunlop J.S., Taylor G.L.,
Hughes D.H., \& Robson E.L., 1993, MNRAS, 264, 445

\bibitem[Dunlop et al. 2001]{dun01} Dunlop J.S., McLure R.J., Kukula M.J., et al., 2001, MNRAS, submitted [astro-ph/0108397]

\bibitem[Eales \& Rawlings 1996]{er96} Eales S.A., \& Rawlings S., 1996,
ApJ, 460, 68

\bibitem[Ellingson et al.\ 1991]{eyg91} Ellingson E., Yee H.K.C., \& 
Green R.F., 1991, ApJ 371, 49

\bibitem[Falomo et al.\ 2001]{fal01} Falomo R., Kotilainen J., \& Treves A., 2001, ApJ, 547, 124

\bibitem[Ferrarese \& Merritt \ 2000]{fm00} Ferrarese L., \& Merritt D., 2000, ApJ, 539, L9

\bibitem[Franceschini et al. 1998]{fra98} Franceschini A., Vercellone S., \& Fabian A.C., 1998,MNRAS, 397, 817

\bibitem[Freeman \ 1970]{free80} Freeman K.C., 1970, ApJ, 160, 812

\bibitem[Gebhardt et al.\ 2000]{geb02} Gebhardt K., Bender, R., Bower, G., et al., 2000, ApJ, 539, L13

\bibitem[Gonz\'alez-Serrano et al. 1993]{gon93} Gonz\'alez-Serrano J.I., Carballo R., \& P\'erez-Fournon I., 1993, AJ, 105, 1710

%\bibitem[Haiman \& Menou 2000]{hm00} Haiman Z., \& Menou K., 2000, ApJ, 531, 42

\bibitem[Hall et al. 1998]{hall98} Hall P.B., Green R.F., \& Cohen M., 1998, ApJS, 119, 1

\bibitem[Hamabe \& Kormendy 1987]{hk87} Hamabe M., \& Kormendy J., {\it Structure and Dynamics of Elliptical Galaxies}, IAU Symp., No.127, p.379, ed. de Zeeuw, T., Reidel, Dordrecht

\bibitem[Heckman et al. \ 1991]{heck91} Heckman T.M., Miley G.K., Lehnert M.D., \& van Breugel W., 1991, ApJ, 370, 78

\bibitem[Hoessel et al. 1987]{hoe87} Hoessel J.G., Oegerle W.R., \& Schneider D.P., 1987, AJ, 94, 1111

\bibitem[Hughes et al. 2000]{hug00} Hughes D.H., Kukula M.J., Dunlop J.S., \& \& Boroson T., 2000, MNRAS, in press

\bibitem[Hutchings \& Neff 1992]{hutn92} Hutchings J.B., Neff S.G.,
1992, AJ, 104,1

\bibitem[Hutchings \& Neff 1997]{hutn97} Hutchings J.B., \& Neff S.G.,
1997, AJ, 113, 550

\bibitem[Hutchings \& Morris 1995]{hutc95} Hutchings J.B.,\& 
Morris S.C., 1995, AJ, 109, 1541

\bibitem[Hutchings et al.\ 1994]{hut84} Hutchings J.B., Holtzman J.,
Sparks W.B., et al., 1994, ApJ, 429, L1

\bibitem[Hutchings et al. 1995]{hut95} Hutchings J.B., 1995, Nature
News\&Views, 376,118

\bibitem[Hutchings et al. 2002]{hut02} Hutchings J.B., Frenette D., Hanisch R., et al., 2002, AJ, 123, 2936

\bibitem[Inskip et al. \ 2002]{in02} Inskip K.J., Best P.N., Longair M.S, \& MacKay D.J.C., 2002, MNRAS, 329, 277

\bibitem[Jarvis et al. \ 2001]{jar01} Jarvis M.J., Rawlings S., Eales S.A., et al., 2001, MNRAS, 326, 1585

\bibitem[Jedrzjewski 1987]{J87} Jedrzejewski R.I., 1987, MNRAS, 226, 747

\bibitem[Kauffmann \& Heaknet 2000]{kh00} Kauffmann G., \& Haehnelt M., 2000, MNRAS, 311, 576

\bibitem[Kirhakos et al. 2001]{kir01} Kirhakos S., Bahcall J.N., Schneider D.P., \& Kristian J., 2001, ApJ, 520, 67

\bibitem[Kolb \& Turner 1990]{kt90} Kolb E.W., \& Turner M.S., 1990, ``The Early Universe'', Addison-Wesley Publishing Company, p. 42-45

\bibitem[Kormendy \& Richstone 1995] Kormendy J., \& Richstone D., 1995, ARA\&A, 33, 581

\bibitem[Kotilainen et al. 1998]{kot98} Kotilainen J.K., Falomo R., \& 
Scarpa R., 1999, AA, 332, 503

\bibitem[Kotilainen et al. 2000]{kot00} Kotilainen J.K., \& Falomo R., 2000,A\&A, 364, 70

\bibitem[Kukula et al. 1997]{kuk97} Kukula M.J., Dunlop J.S.,
Hughes D.H., Taylor G., \& Boroson T., 1997, Quasar Hosts, Ed. D.L.Clements \&
  I.P\'erez-Fourn\'on, ESO-IAC Conference, pg. 177

\bibitem[Kukula et al. 2001]{kuk01} Kukula M.J., Dunlop J.S., McLure R.J., Miller L., et al., 2001, MNRAS, 325, 1533

\bibitem[Lacy et al. 2000]{lay00} Lacy M., Bunker A.J., \& Ridgway S.E., 2000, AJ, 120, 68

\bibitem[Ledlow \& Owen 1996]{led96} Ledlow M.J., \& Owen F.N., 1996, AJ, 112, 9

\bibitem[Lee \& Hwang \ 2000]{lh00} Lee M.G., \& Hwang N., 2000, ISASS, 14, 131

\bibitem[Lehnert et al. 1992]{len92} Lehnert M.D., Heckman T.M., Chambers K.C., \& Miley G.K., 1992, ApJ, 393, 68

\bibitem[Lehnert et al. 1999a]{len99a} Lehnert M., Miley G.K.,
Sparks W.B., et al., 1999, ApJSS, 123, 351

\bibitem[Lehnert et al. 1999b]{len99b} Lehnert M., Matthew D., van
Breugel W.J.M., Heckman T.M., \& Miley G.K., 1999, ApJSS, 124, 11

\bibitem[Lilly 1989]{lil89} Lilly S.J., 1989, ApJ, 340, 77

\bibitem[Lilly et al. 1985]{lil85} Lilly S.J., Longair M.S., \& 
  Allington-Smith J.R., 1985, MNRAS, 215, 37

\bibitem[MacKenty 1990]{mac90} MacKenty J.W., 1990, ApJS, 72, 231

\bibitem[Magorrian et al. 1998]{mag98} Magorrian J.
Tremaine S.,Richstone D., et al., 1998, AJ, 115, 2285

\bibitem[McLeod \& Rieke 1994a]{mcl94a} McLeod K.K., \& Rieke G.H., 1994, ApJ, 420, 58

\bibitem[McLeod \& Rieke 1994b]{mcl94b} McLeod K.K., \& Rieke G.H., 1994, ApJ, 431, 137

\bibitem[McLeod \& Rieke 1995]{mcl95} McLeod K.K., \& Rieke G.H., 1995, ApJ, 454, L77

%\bibitem[McLeod et al. 1999]{mcl99} McLeod K.K., Rieke G.H., \& Storrie-Lombardi L.J., 1999, ApJ, 511, L67

\bibitem[McLeod \& McLeod 2001]{mcl01} McLeod K.K., \& McLeod B.A., 2001, ApJ, 546, 782

\bibitem[McLure et al. 1999]{mc99} McLure R.J., Kukula M.J., Dunlop J.S., et al., 
1999, MNRAS, 308, 377

\bibitem[McLure \& Dunlop 2002]{mc02} McLure R.J., \& Dunlop J.S., 2002, MNRAS, 331, 795

\bibitem[Meurer et al 1995]{meu95} Meurer G. R., Heckman T. M., Leitherer C., et al., 
  1995, ApJ, 110, 2665

\bibitem[Moffat 1969]{mof69} Moffat A.F.J., 1969, A\&A, 3, 455

\bibitem[Mobasher et al. 1993]{mob93} Mobasher B., Sharples R.M., \&
Ellis R.S., 1993, MNRAS, 263, 560

\bibitem[Mobasher et al. 1999]{mob99} Mobasher B., Guzm\'an R., Arag\'on-Salamanca A., \& Zepf S., 1999, MNRAS, 304, 225

\bibitem[Nolan et al. 2001]{no01} Nolan L.A., Dunlop J.S., Kukula M.J., et al., 2001, MNRAS, 323, 308

\bibitem[Percival et al. 2001]{per00} Percival W.J., Miller L., McLure R.J., \& Dunlop J.S., 2001, MNRAS, 322, 843

\bibitem[Rawlings et al. 1991]{raw91} Rawlings S., \& Saunders R., 1991, Nature, 349, 138

\bibitem[Rawlings et al. 1998]{raw98} Rawlings S., Blundell K.M.,
Lacy M., Willot C.J., \& Eales S.A., 1998,   Proceedings of the workshop
``Observational Cosmology with Radio Surveys'', Kluwer, ASSL, vol 226,
171 [astro-ph/9704151]

\bibitem[Ridgway \& Stockton 1997]{RS97} Ridgway S.E., \& Stockton A.,
1997, AJ,  114, 511

\bibitem[Ridgway et al. \ 2001]{RS01} Ridgway S.E., Heckman T.M., Calzetti D., \& Lehnert M., 2001, ApJ, 550, 122

\bibitem[Riegler et al. 1992]{rieg92} Riegler M.A., Lilly S.J.,
Stockton A., Hammer F., \& Le F\`evre 0., 1992, ApJ, 385, 61

\bibitem[Salpeter \ 1955]{sal55} Salpeter E.E., 1955, ApJ, 121, 161

\bibitem[S\'anchez 2001]{yo01} S\'anchez S.F., 2001, PhD Thesis, Univ. of Cantabria.

\bibitem[S\'anchez \& Gonzalez-Serrano 1999]{san99} S\'anchez S.F., \& Gonzalez-Serrano J.I., 1999, A\&A, 352, 395

\bibitem[S\'anchez \& Gonzalez-Serrano 2002]{san02} S\'anchez S.F., \& 
  Gonzalez-Serrano J.I., 2002, A\&A, 396, 773

\bibitem[Serjeant et al. 1998]{ser98} Serjeant S., Rawlings S., Maddox S.J., et al., 1998, MNRAS, 294, 494

\bibitem[Shapley \& Ames \ 1932]{SAC32} Shapley H., \& Ames A., 1932, Harvard Ann. Vol. 88, No. 2

\bibitem[Simpson \& Rawlings 2000]{SR00} Simpson C., \& Rawlings S., 2000, MNRAS, 317, 1023

\bibitem[Smith \& Heckman 1990]{SH90} Smith E.P., \& Heckman T.M., 1990,
ApJ, 348,  38

\bibitem[Smith et al.\ 1986]{sm86} Smith E.P., Heckman T.M., Bothun G.D.
, Romanishin W., \& Balick B., 1986, ApJ 306, 64

\bibitem[Stockton \& Ridgway 2001]{RS01} Stockton A., \& Ridgway S.E., 
2001, 554, 1012

\bibitem[Taylor et al. 1996]{tay96} Taylor G.T., Dunlop J.S.,
Hughes D.H., \& Robson E.I., 1996, MNRAS, 283, 930

\bibitem[Urry \& Padovani 1995]{up95} Urry C.M., \& Padovani P., 1995,
PASP, 107, 803

\bibitem[van den Bergh \ 1996]{vdb96} van den Bergh S., Abraham R.G., Ellis R.S., Tanvir N.R., \& Glazebrook K.G., 1996, AJ, 112, 359

\bibitem[Van Breugel et al. 1999]{van99} Van Breugel W., de Breuck C.,
  Stanford S.A., Stern D., R\"ottgering H., \& Miley G., 1999, ApJ, 518, 61

\bibitem[V\'eron-Cetty \& Wotjer 1990]{vcw90} V\'eron-Cetty M.P., \& 
Woltjer L., 1990, AA, 236, 69

\bibitem[Vigotti et al. \ 1997]{vig97} Vigotti M., Vettolani G.V.,
Merighi R., Lahulla J.F., \& Pedani M., 1997, A\&AS, 123, 1

\bibitem[whitmore et al. \ 1999]{whi99} Whitmore, B.C, Zhang Q., Leitherer C., et al., 
  1999, ApJ, 118, 1551

\bibitem[Willot et al. \ 1999]{W99} Willot C.J., Rawlings S., Blundell K.M., \& Lacy M., 1999, MNRAS, 309, 1017

\bibitem[Willot et al. \ 2001]{W01} Willot C.J., Rawlings S., \& Blundell K.M., 2001, MNRAS, 324, 1

\bibitem[Willot et al. \ 2002]{W02} Willot C.J., Rawlings S., Jarvis M.J., \& Blundell K.M., 2002, MNRAS, in press [astro-ph/0209439]

\end{thebibliography}
\end{document}